\documentclass[11pt,english]{article}
\usepackage{lmodern}

\usepackage[T1]{fontenc}
\usepackage[latin9]{inputenc}
\usepackage{geometry}
\usepackage{float}
\usepackage{amsmath}
\usepackage{amsthm}
\usepackage{amssymb}
\usepackage{graphicx}
\usepackage{setspace}
\usepackage[authoryear]{natbib}
\makeatletter

\providecommand{\tabularnewline}{\\}

\setcitestyle{round}
\usepackage{lscape}
\usepackage{longtable}
\usepackage{graphicx}
\usepackage{tabularx,subfig}

\makeatother
\usepackage{babel}

\begin{document}
\title{Relative Contagiousness of Emerging Virus Variants:\\
 An Analysis of the Alpha, Delta, and Omicron SARS-CoV-2 Variants
 \thanks{\small{}Thanks to Torben Andersen, Victor Chernozhukov, Peter Dalgaard, Emily Dyckman, 
Claus Ekstr{\o}m, Mogens Fosgerau, Ulrik Gerdes, Martin Vin{\ae}s Larsen, Serena Ng, 
Uffe Poulsen, Morten Rasmussen, and Tom Wenseleers for valuable comments and 
suggestions. I also thank anonymous reviewers for valuable comments and suggestions 
and David Sanders and Michael Krabbe Borregaard for making tutorials and information
about the Julia language available and the Danish Patient Safety Authority
for sharing data on COVID-19 cases in relation to Euro 2020 games.} }
\medskip{}

\author{\textbf{Peter Reinhard Hansen}$^{a,b}$\bigskip{}
\\
{\normalsize{}$^{a}$}\emph{\normalsize{}University of North Carolina}\textbf{}\thanks{Address: University of North Carolina, Department of Economics, 107
Gardner Hall Chapel Hill, NC 27599-3305}\emph{\normalsize{}
}\\
{\normalsize{}$^{b}$}\emph{\normalsize{}Copenhagen Business School\smallskip{}
}}
\date{\emph{\normalsize{}\today}}
\maketitle
\begin{abstract}
{\small{}We propose a simple dynamic model for estimating the
relative contagiousness of two virus variants. Maximum likelihood
estimation and inference is conveniently invariant to variation in
the total number of cases over the sample period and can be expressed
as a logistic regression. 
We apply the model to Danish SARS-CoV-2 variant data. We estimate the reproduction numbers of Alpha and Delta to be larger than that of the ancestral variant by a factor of 1.51 {[}CI 95\%: 1.50, 1.53{]} and 3.28 {[}CI 95\%: 3.01, 3.58{]}, respectively. In a predominately vaccinated population, we estimate Omicron to be 3.15 {[}CI 95\%: 2.83, 3.50{]} times more infectious than Delta.

Forecasting the proportion of an emerging virus variant is
straight forward and we proceed to show how the effective reproduction
number for a new variant can be estimated without contemporary sequencing
results. This is useful for assessing the state of the pandemic in
real time as we illustrate empirically with the inferred effective
reproduction number for the Alpha variant.}{\normalsize\par}
\end{abstract}
\textit{\small{}Keywords:}{\small{}COVID-19, SARS-CoV-2, Reproduction
number, Alpha variant, Delta variant, Omicron variant, B.1.1.7, B.1.617.2, B.1.1.529, Maximum
Likelihood, Logistic Regression.}{\small\par}\medskip

\newpage

\section{Introduction}

During the fall of 2020, confirmed cases of COVID-19 grew rapidly
in the UK with the emergence of the Alpha variant of SARS-CoV-2 (B.1.1.7)
formerly known as the British variant, see \citet{B117found}. The
Alpha variant was shown to be more contagious than earlier lineages,
see \citet{Volz2020.12.30.20249034} and \citet{Washington2021.02.06.21251159}.
Moreover, infection with the Alpha variant was found to increase the
risk of hospitalization by about 42\%, see \citet{BagerEtAl2021}.
India experienced a similar explosive growth in COVID-19 cases in
April 2021 following the emergence of the Delta variant  (B.1.617.2),
formerly known as the Indian variant. The Omicron variant (B.1.1.529) was first detected on November 22, 2021 
in southern Africa, see \citet{Omicron}, and has led to rapid increases in COVID-19 case numbers in many parts of the world.

In this paper, we formulate a simple model for two virus variants
of an infections disease, where the object of interest is the relative
contagiousness, denoted by $\gamma$, which we will also refer to as the competitive advantage. The time series of new-variant
cases to total cases can be modeled as binomially distributed variables
with a time-varying parameter, $\lambda_{t}$. The dynamic properties
of $\lambda_{t}$ are given by the relative contagiousness that can
be estimated solely from changes in the relative proportion of the
two variant. The analysis is therefore invariant to the reproduction
number for the existing lineage and time variation therein. This is
convenient because the reproduction number varies substantially over
time due to changes in behavior and preventive measures along with
other factors. The analysis is also invariant to testing intensity
and time variation therein, so long as the variation in sampling does
not ``favor'' one variant over another. Our starting point is maximum
likelihood analysis of the sequenced tests, which leads to a simple
logistic regression after some straight forward algebra. This greatly
simplifies the estimation, inference, and prediction.\footnote{A early draft with these results were first shared on January 26, 2021, https://twitter.com/ProfPHansen/status/1353808634652708864 and a YouTube presentation of these results and an analysis of the Alpha variant (B.1.1.7) was shared on February 14, 2021. https://www.youtube.com/watch?v=u5kkwe3aXLM}

The \emph{basic reproduction number}, $R_{0}$, is an important characteristic
of an infections disease and during a pandemic the \emph{effective
reproduction number,} $R_{t}$, provides a measure of the direction
for case numbers during the pandemic. The time it takes to determine
the genome in positive cases present an obstacle for assessing the
reproduction number for a new virus variant. Such data are typically
only available with some delay. We show that the highly predictable
nature of the proportion of a new variant can be used to infer its
reproduction number from the aggregate reproduction number and the
most recently available estimate of the proportion, $\lambda$. The
effective reproduction number for all cases is simpler to compute
and the proportion of a new variant can be projected forward, typically
with high accuracy. This makes it possible to compute the effective
reproduction number of a new variant with a simple formula before
contemporaneous sequencing results are available.

\begin{figure}[h]
    \centering
        \includegraphics[height=0.3\textheight]{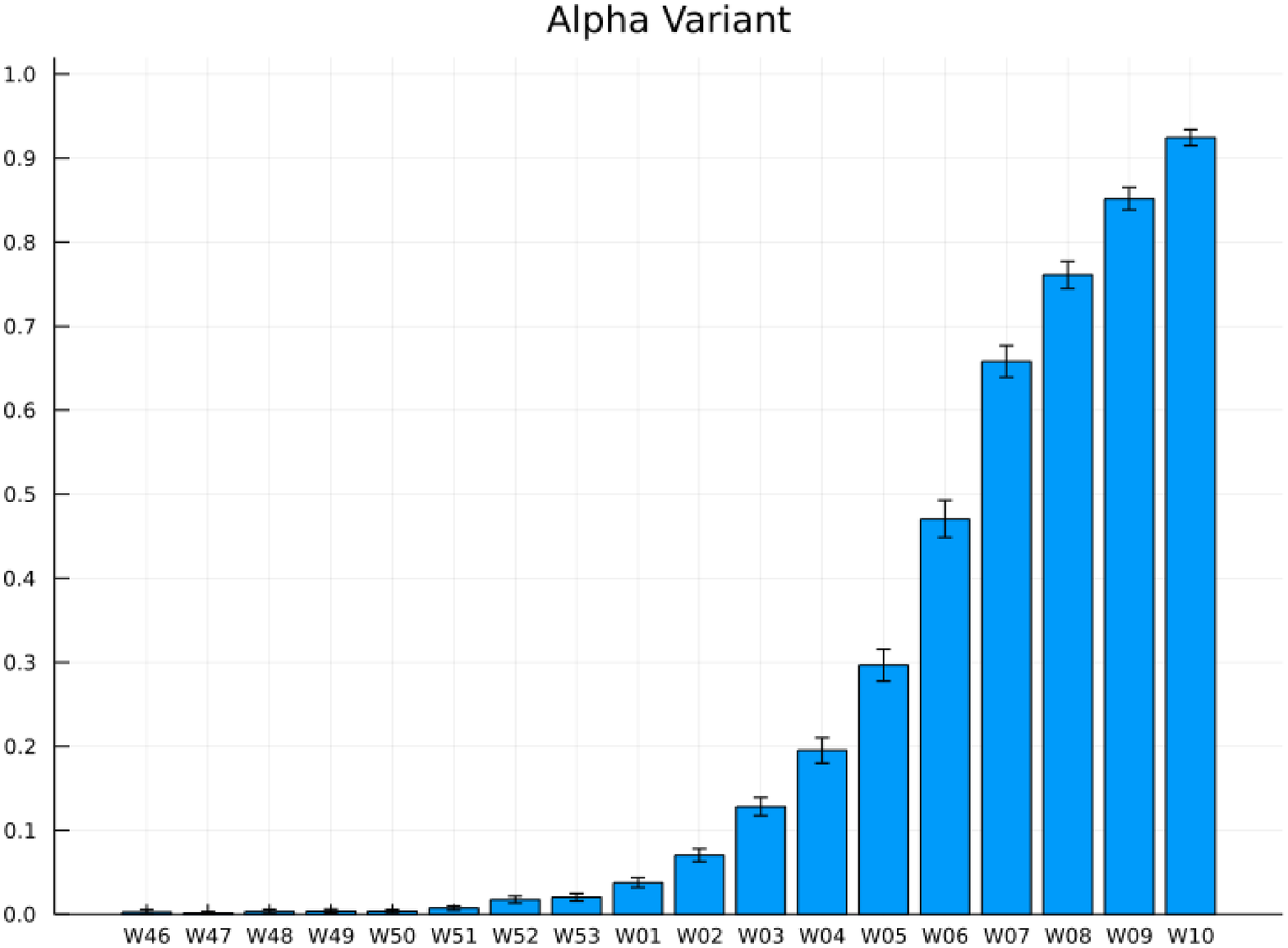}%
        \includegraphics[viewport=0bp 3bp 450bp 550bp,height=0.304\textheight]{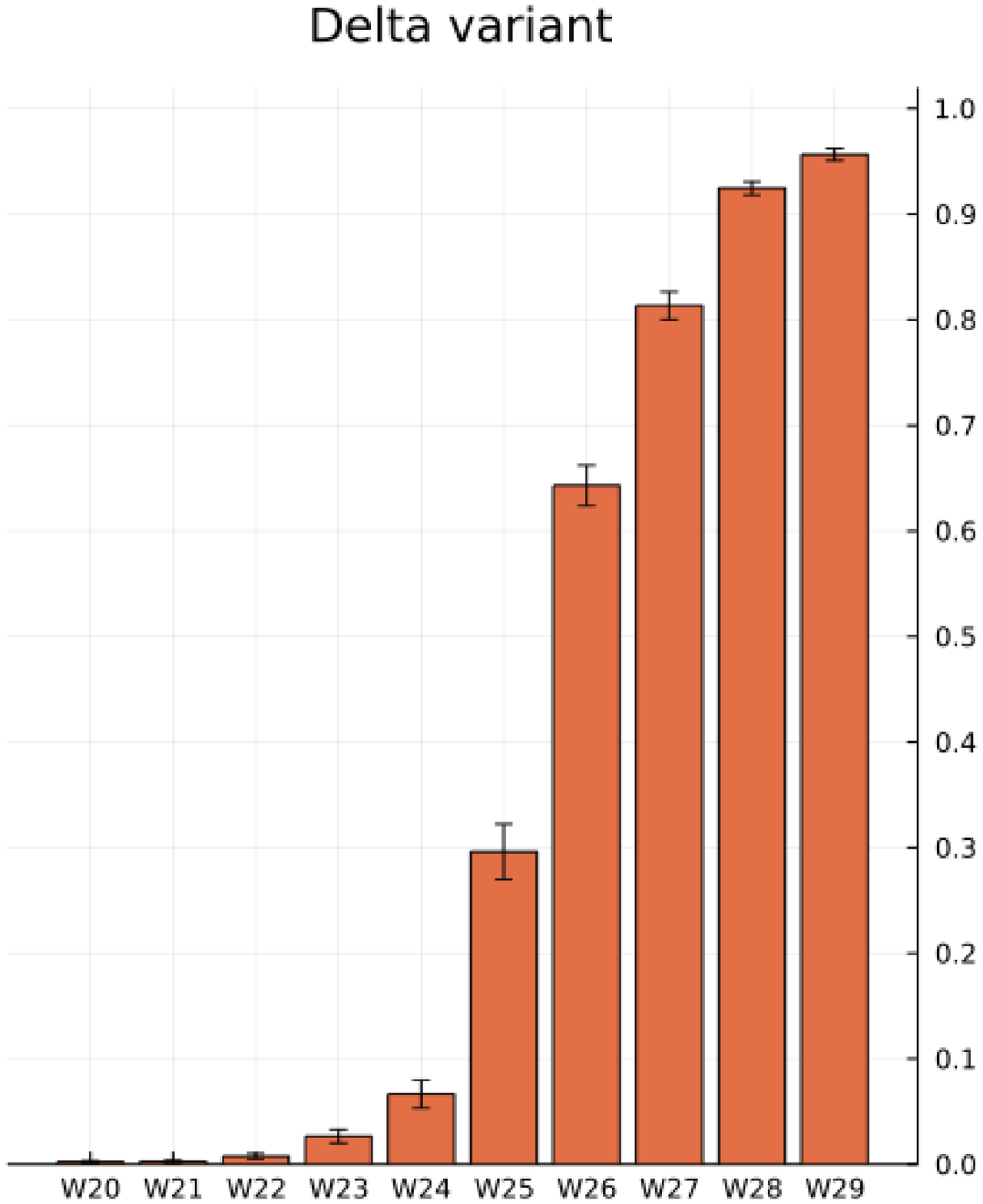}%
        \caption{Weekly proportions of Alpha and Delta variants relative to all cases.\label{fig:ProportionAlphaAndDelta}}
\end{figure}

We apply the methodology to Danish data and study the Alpha and Delta variants using weekly data, whereas daily data are used to study the swiftly progressing Omicron variant. The Danish
data are excellent for studying the progression of a new variant of
SARS-COV-2 because the vast majority of confirmed COVID-19 cases were sequenced during the periods where each of the three variants emerged. 
Moreover, testing is extensive in Denmark. During the three sample periods, the number
of weekly PCR tests varied between nearly 500 thousands and over 1.5
million tests per week for a population of about 5.8 million individuals. The proportion of the new-variant
cases increased from $<$0.5\% to over $90\%$ during the
sample periods for the Alpha and Delta variants, see Figure \ref{fig:ProportionAlphaAndDelta}, and from  $<$2\% to over $90\%$ during the sample period for the Omicron variant.
The progression of the Alpha variant is shown in the left panel Figure
\ref{fig:ProportionAlphaAndDelta} and the progression of the Delta
variant in the right panel, along with 95\% confidence intervals for
each week. It can be seen that the Delta variant progressed substantially
faster than the Alpha variant. For instance, it took the Alpha variant
about eight weeks to increase from $<$10\% to $>$90\%, while this same
growth was achieved by the Delta variant in just 4 weeks. The Omicron variant progressed even faster as we show below. On December 8th, 2021, just 16 days after Omicron was first detected in southern Africa, the Omicron variant accounted for over 10\% of COVID-19 cases in Denmark and 19 days later, on December 28, 2021, the new variant accounted for 90\% of all COVID19 cases, see Table \ref{tab:Daily-Danish-data}.

The paper is organized as follows: We present the statistical model
in Section 2, where the time series of binomially distributed variables
is expressed as a logistic regression. We present the data and the
empirical results on relative contagiousness in Section 3. Section
4 presents the two auxiliary results on prediction and inferring the
latent reproduction number for a new emerging variant. We illustrate
how the proportion of new-variant cases can be predicted and derive
the associated confidence bands. We apply this to the Alpha variant
data and review how accurate the estimated model was at predicting
the realized proportion of the Alpha variant out-of-sample. Then we
derive the method for estimating the latent reproduction number for
a new emerging variant and apply the method to the Alpha variant data.
Section 5 has some concluding remarks, and we present extensions and some details
about the maximum likelihood estimation and robust standard errors
in the Appendix B.

\section{The Statistical Model}

In this section we present the simple dynamic structure for the case
numbers of two competing virus variants. The structure is not specific
to competing virus variants, but could be used to analyze other competing
objects. The extension to the situation with more that two competing variants is derived in Appendix A.

\subsection{Two Competing Virus Variants}

Consider a virus with two variants, $A$ and $B$, and let $R_{0}^{A}$
and $R_{0}^{B}$ denote their basic reproduction numbers. We use $B$
to represent a new, emerging variant whereas $A$ represents the older
variant. The parameter of interest is the relative contagiousness,
defined by $\gamma=R_{0}^{B}/R_{0}^{A}$.

Suppose that the number of new cases in period $t$ are denoted $A_{t}$
and $B_{t}$, respectively. The rate of growth in case numbers for
the old variant is denoted, $a_{t}=A_{t}/A_{t-1}$, which depends
on its contagiousness and the number of ``opportunities'' the virus
has to jump from an infected individual to another person. The latter
is heavily influenced by preventive measures and restrictions imposed
by health authorities, seasonality, percentage of susceptible people
in the population, individual behavior, along with many other things.
The new virus variant is subject to the same level of ``opportunities'',
and only differs in terms of its contagiousness, such that the ratio
of effective reproduction numbers is constant and equal to, $\gamma=R_{t}^{B}/R_{t}^{A}$.
It follows that variant $B$'s rate of growth is proportional to $a_{t}$,
and given by
\[
b_{t}=B_{t}/B_{t-1}=\gamma a_{t}.
\]

\begin{figure}
    \centering
        \includegraphics[width=0.49\textwidth]{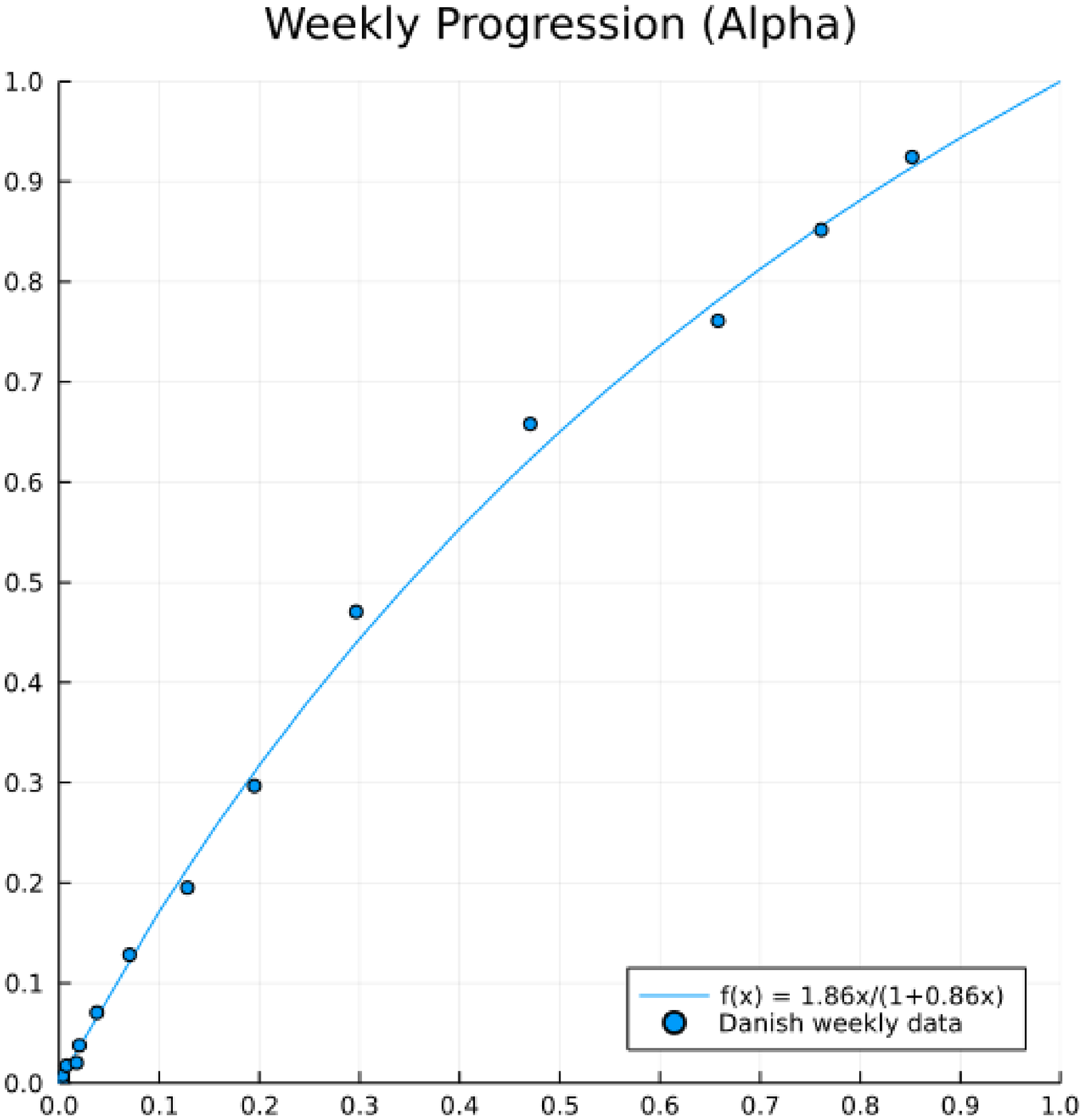}%
        \includegraphics[width=0.49\textwidth]{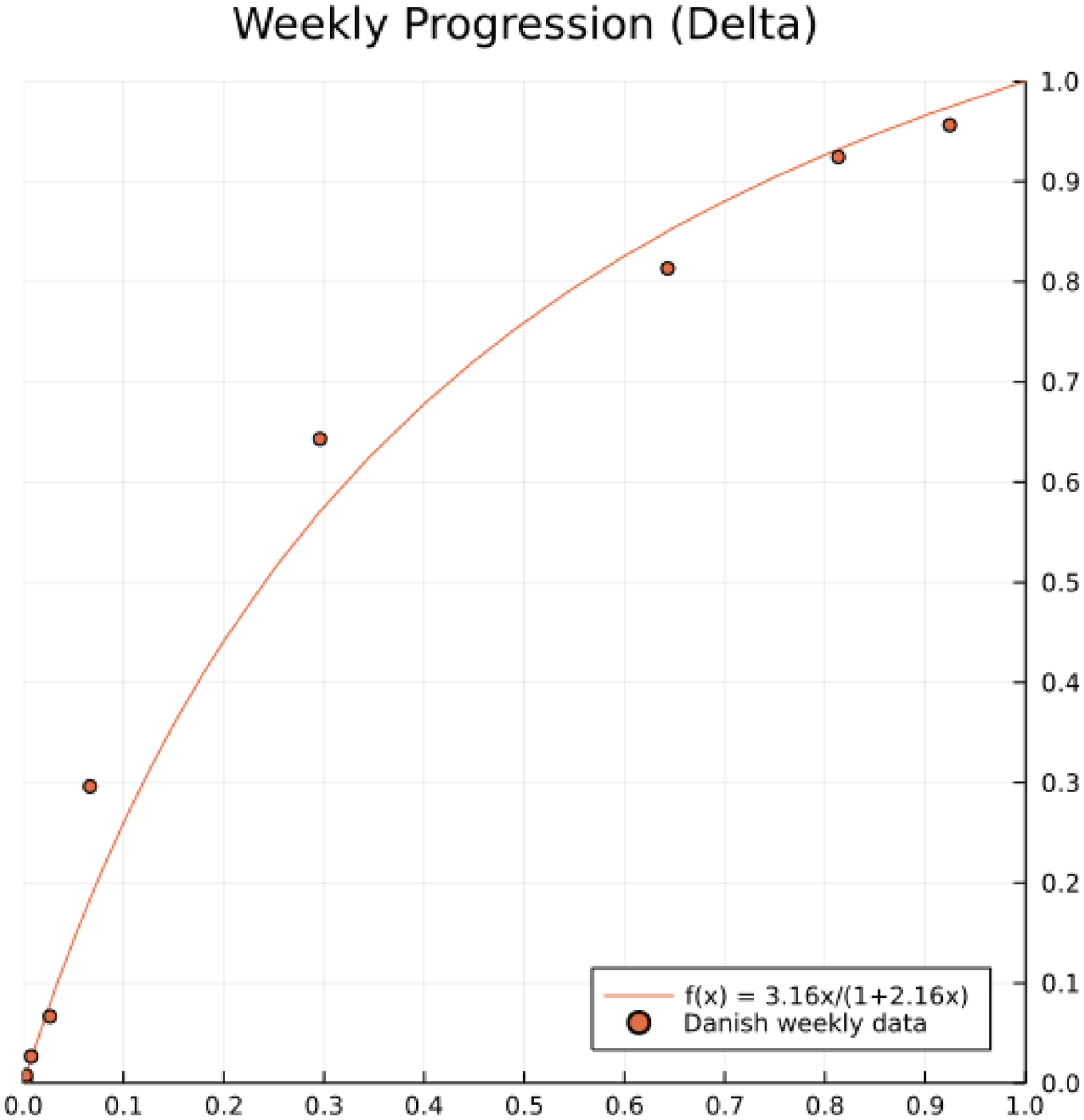}%
        \caption{Proportion of the new variant 
            against that of the previous week.\label{fig:ProportionScatter}}
\end{figure}

The observed data are the detected cases for which the genome is determined.
In period $t$, the genome is determined in $N_{t}$ of the new cases,
where $X_{t}$ are variant $B$ and $N_{t}-X_{t}$ are variant $A$.
We assume $N_{t}$ is a representative random sample from the population
of new cases such that $X_{t}\sim\mathrm{Bin}(N_{t},\lambda_{t})$,
with $\lambda_{t}=\frac{B_{t}}{A_{t}+B_{t}}$. It follows that 
\begin{eqnarray}
\lambda_{t+1}=\frac{B_{t+1}}{A_{t+1}+B_{t+1}} & = & \frac{\gamma a_{t+1}B_{t}}{a_{t+1}A_{t}+\gamma a_{t+1}B_{t}}\nonumber \\
 & = & \frac{\gamma B_{t}/(A_{t}+B_{t})}{A_{t}/(A_{t}+B_{t})+\gamma B_{t}/(A_{t}+B_{t})}=\frac{\gamma\lambda_{t}}{(1-\lambda_{t})+\gamma\lambda_{t}}.\label{eq:lambda}
\end{eqnarray}
Note that the dynamic equation for $\lambda_{t}$ depends on the ratio
$\gamma=b_{t}/a_{t}$ but not the actual values of $a_{t}$ and $b_{t}$.
This greatly simplifies the analysis and estimation and inference
about $\gamma$ becomes invariant to a range of changes during the
sample period that influence the rate of change in case numbers. Equation
(\ref{eq:lambda}) defines the function $f(x)=\gamma x/[1+(\gamma-1)x]$ that characterizes the expected progression of a new variant. Figure \ref{fig:ProportionScatter}
presents two examples of $f(x)$, which
 is strictly increasing for $\gamma>1$. The case $\gamma=1.86$ is shown
in the left panel and $\gamma=3.16$ in the right panel along with
the progression of the weekly observed progression of the Alpha (left) and Delta (right) variants, where the proportion of the new variant this week (y-axis) is plotted against its proportion in the previous week (x-axis).

The progression of a very contagious variant can seem deceptively
slow in the early phase. If we take the case $\gamma=1.86$, which
is our estimate for the Alpha variant, then it takes four weeks for
the new variant to increase from 1 in 1,000 cases to 1 in a 100 cases.
Then another four weeks to increase to 1 in 10 cases. After that,
it picks up the pace and the new variant becomes the dominant variant
($\lambda_{t}>50\%$) four weeks later and reaches +90\% of all cases
after another four weeks. So, it can take months from the
moment the first case of a new and more contagious variant is observed
to the time when the new variant begins to have a noticeable impact
on the total number of cases. After the new variant becomes dominant
it can radically change the momentum of the pandemic. A relatively
stable period can suddenly become one where cases grow exponentially,
because the new dominant variant has a larger reproduction number.
This scenario played out for both the Alpha, Delta, and Omicron variants in
many places, such as the August 2021 large surge in Delta cases in Florida,
Texas, along with other states in the US.

\subsection{The Likelihood Analysis and Logistic Regression}

Let $N_{t}$ denote the total number of new cases in period $t$ for
which the genome is identified, and let $X_{t}$ be the number of
cases that are identified as the new emerging variant. We assume that $X_t$ is binomially distributed, $\mathrm{Bin}(N_t,\lambda_t)$, where $\lambda_t$ evolves according to (\ref{eq:lambda}). This requires that the variant is identified in a representative sample of positive cases for all $t$. In the absence of serial dependence, the log-likelihood
function for the sample $(N_{t},X_{t})$, $t=1,\ldots,T$, is proportional
to 
\[
\ell(\gamma,\lambda_{0})\propto\sum_{t=1}^{T}X_{t}\log\lambda_{t}+(N_{t}-X_{t})\log(1-\lambda_{t}).
\]
The
two unknown parameters, the initial value $\lambda_{0}\equiv\lambda$
and $\gamma$, can be estimated by maximum likelihood, $(\hat{\lambda},\hat{\gamma})=\arg\max_{\gamma,\lambda}\ell(\lambda,\gamma)$,
and confidence intervals for $\lambda$ and $\gamma$ can be obtained
with conventional methods. The likelihood can conveniently be expressed
as a logistic regression model. For this purpose, we introduce the
odds ratio, $\rho_{t}=\lambda_{t}/(1-\lambda_{t})$, and it is simple
to show that (\ref{eq:lambda}) is equivalent to the simple dynamic
equation, $\rho_{t}=\gamma\rho_{t-1}$. This implies that 
\[
\rho_{t}=\gamma^{t}\rho_{0}=\exp(\log\rho_{0}+\log\gamma\times t)=\exp(\alpha+\beta t),
\]
where $\alpha=\log\rho_{0}$ and $\beta=\log\gamma$. Since $\lambda_{t}=\rho_{t}/(1+\rho_{t})$,
the structure of the logistic regression model emerges such that
\begin{equation}
\lambda_{t}=\frac{\exp\left(\alpha+\beta t\right)}{1+\exp\left(\alpha+\beta t\right)}=\frac{1}{1+e^{-\alpha-\beta t}}.\label{eq:LogisticReg}
\end{equation}
This model is straight forward to estimate and analyze using standard
software implementations, including the generalized linear model package,
\texttt{glm}, that is implemented in R and Julia. In the empirical
analysis we estimate the model by maximum likelihood and compute robust
standard errors from the score and hessian of the log-likelihood function,
see \citet{White94}. While heteroskedasticity is a natural feature
of binomially distributed random variables with a time-varying probability,
$\lambda_{t}$, distributional misspecification can invalidate the
non-robust standard errors. Dynamic misspecification where $\lambda_{t}=(1+e^{-\alpha-\beta t})^{-1}$
does not hold for all $t$ can result in inconsistent estimates. The
details are presented in Appendix B.\footnote{Identical estimates were obtained with the \texttt{glm} packaged in
Julia, see \citet{2019arXiv190708611B} and \citet{dahua_lin_2021_5105997}.
The proper command for the \texttt{glm} package in Julia is: \texttt{glm(@formula(x
/ n \textasciitilde{} time\_trend), {[}data{]}, wts = n, Binomial())}
and in $\mathtt{R}$ it is: \texttt{glm(x/n \textasciitilde{} tt,
weights=n, {[}data{]}, family = binomial)}, see \citet{Rmanual} for
details. The latter was kindly provided by Peter Dalgaard. The \texttt{glm}
package computes the non-robust standard errors based on the Fisher
information. These were smaller than the robust standard errors, in
particular in our analysis of the Delta variant. Robust and non-robust
confidence intervals are reported in Appendix B.} 
\begin{table}[!ht]
\caption{COVID-19 PCR tests and outcomes by week} 
\begin{center}
\begin{tabular*}{\textwidth}{cccccc}
\hline\hline 
Week & Tested & Cases & Sequenced & Alpha cases & Alpha proportion\\
 & (PCR) & $C_{t}$ & $N_{t}$ ($N_{t}/C_{t}$) & $X_{t}$ & $X_{t}/N_{t}$\\
\hline 
\hline 
46 & 490,543 & 7,533 & 1,486 (19.7\%) & 4 & 0.27\%\tabularnewline
47 & 502,852 & 8,456 & 1,941 (23.0\%) & 3 & 0.15\%\tabularnewline
48 & 502,851 & 8,774 & 2,127 (24.2\%) & 7 & 0.33\%\tabularnewline
49 & 544,578 & 12,816 & 2,868 (22.4\%) & 11 & 0.38\%\tabularnewline
50 & 694,989 & 21,925 & 4,226 (19.3\%) & 16 & 0.38\%\tabularnewline
51 & 883,253 & 24,579 & 4,943 (20.1\%) & 37 & 0.75\%\tabularnewline
52 & 650,374 & 17,043 & 3,633 (21.3\%) & 64 & 1.76\%\tabularnewline
53 & 536,958 & 14,560 & 3,916 (26.9\%) & 80 & 2.04\%\tabularnewline
1 & 563,348 & 11,311 & 4,161 (36.8\%) & 157 & 3.77\%\tabularnewline
2 & 596,048 & 7,008 & 4,230 (60.4\%) & 298 & 7.04\%\tabularnewline
3 & 739,922 & 5,321 & 3,688 (69.3\%) & 473 & 12.83\%\tabularnewline
4 & 768,925 & 3,616 & 2,660 (73.6\%) & 519 & 19.51\%\tabularnewline
5 & 794,917 & 3,096 & 2,235 (72.2\%) & 663 & 29.66\%\tabularnewline
6 & 809,028 & 2,716 & 1,974 (72.7\%) & 929 & 47.06\%\tabularnewline
7 & 833,795 & 3,335 & 2,416 (72.4\%) & 1,590 & 65.81\%\tabularnewline
8 & 956,070 & 3,688 & 2,683 (72.7\%) & 2,042 & 76.11\%\tabularnewline
9 & 1,033,111 & 3,616 & 2,699 (74.6\%) & 2,299 & 85.18\%\tabularnewline
10 & 1,056,404 & 3,809 & 2,874 (75.5\%) & 2,657 & 92.45\%\tabularnewline
\hline 
 &  &  &  &  & \tabularnewline
Week & $\quad$Tested$\quad$ & $\quad$Cases$\quad$ & Sequenced & Delta cases & Delta proportion\tabularnewline
 & (PCR) & $C_{t}$ & $N_{t}$ ($N_{t}/C_{t}$) & $X_{t}$ & $X_{t}/N_{t}$\tabularnewline
\hline 
20 & 1,167,981 & 6,867 & 5,366 (78.1\%) & 13 & 0.24\%\tabularnewline
21 & 1,013,403 & 6,698 & 5,213 (77.8\%) & 15 & 0.29\%\tabularnewline
22 & 911,764 & 5,662 & 4,565 (80.6\%) & 36 & 0.79\%\tabularnewline
23 & 720,274 & 2,811 & 2,467 (87.8\%) & 66 & 2.68\%\tabularnewline
24 & 575,207 & 1,649 & 1,364 (82.7\%) & 91 & 6.67\%\tabularnewline
25 & 524,837 & 1,315 & 1,165 (88.6\%) & 345 & 29.61\%\tabularnewline
26 & 608,540 & 2,674 & 2,418 (90.4\%) & 1,555 & 64.31\%\tabularnewline
27 & 624,414 & 4,614 & 3,322 (72.0\%) & 2,702 & 81.34\%\tabularnewline
28 & 583,932 & 6,818 & 6,253 (91.7\%) & 5,781 & 92.45\%\\
29 & 473,843 & 5,289 & 4,800 (90.8\%) & 4,591 & 95.65\% \\
\hline\hline 
\end{tabular*}
\end{center}
\footnotesize
     \renewcommand{\baselineskip}{11pt}
     \textbf{Note:} Source is \emph{Status for udvikling af B.1.1.7 og
andre mere smitsomme varianter i Danmark}, SSI, April 7, 2021 and
\emph{Status for udvikling af SARS-CoV-2 Varianter der overv{\aa}ges i
Danmark} SSI, August 27, 2021. Data available at: https://files.ssi.dk/covid19/virusvarianter/status/status-virusvarianter-07042021-dg45
and https://files.ssi.dk/covid19/virusvarianter/status/virusvarianter-covid-19-280721-gd14\label{tab:Weekly-Danish-data}
\end{table}

\section{Empirical Analyses of the Alpha, Delta, and Omicron Variants}\label{sec:Empirical-Analysis}

Data for the sequenced COVID-19 tests were obtained from the
Statens Serum Institut (SSI), Denmark. We use the weekly data to analyze the Alpha and Delta variants and daily data to analyze the Omicron variant.\footnote{Replication files (Jypyter notebooks with data and Julia code) are available at github.com/reinhardhansen/VirusVariantsReplicationFiles, see \citet{Hansen:2022}.}

The majority of positive
COVID-19 tests had their genome identified in Denmark during the three sample periods, with the exception of the last third of the daily Omicron data, where SSI determined the variant in a representative sample of positive PCR tests.

\subsection{Alpha and Delta Variants}

Table \ref{tab:Weekly-Danish-data} presents the the weekly numbers of PCR COVID-19 tests, the number of positive tests, $C_t$, the number of tests with the genome identified, $N_{t}$, and the number of tests for which the
new emerging variant was found, $X_{t}$, 
along with the percentages of positive tests for which the genome
was determined and the percentage of these tests that were the new
variant.

The first sample period, November 9, 2020 to March 14, 2021 (18 weeks),
is the period where the Alpha variant made its inroad in Denmark,
the second sample period, May 17 to July 25, 2021 (10 weeks), is the
period where the Delta variant grew to dominance in Denmark.

\begin{figure}[h]
\begin{centering}
\includegraphics[height=0.3\textheight]{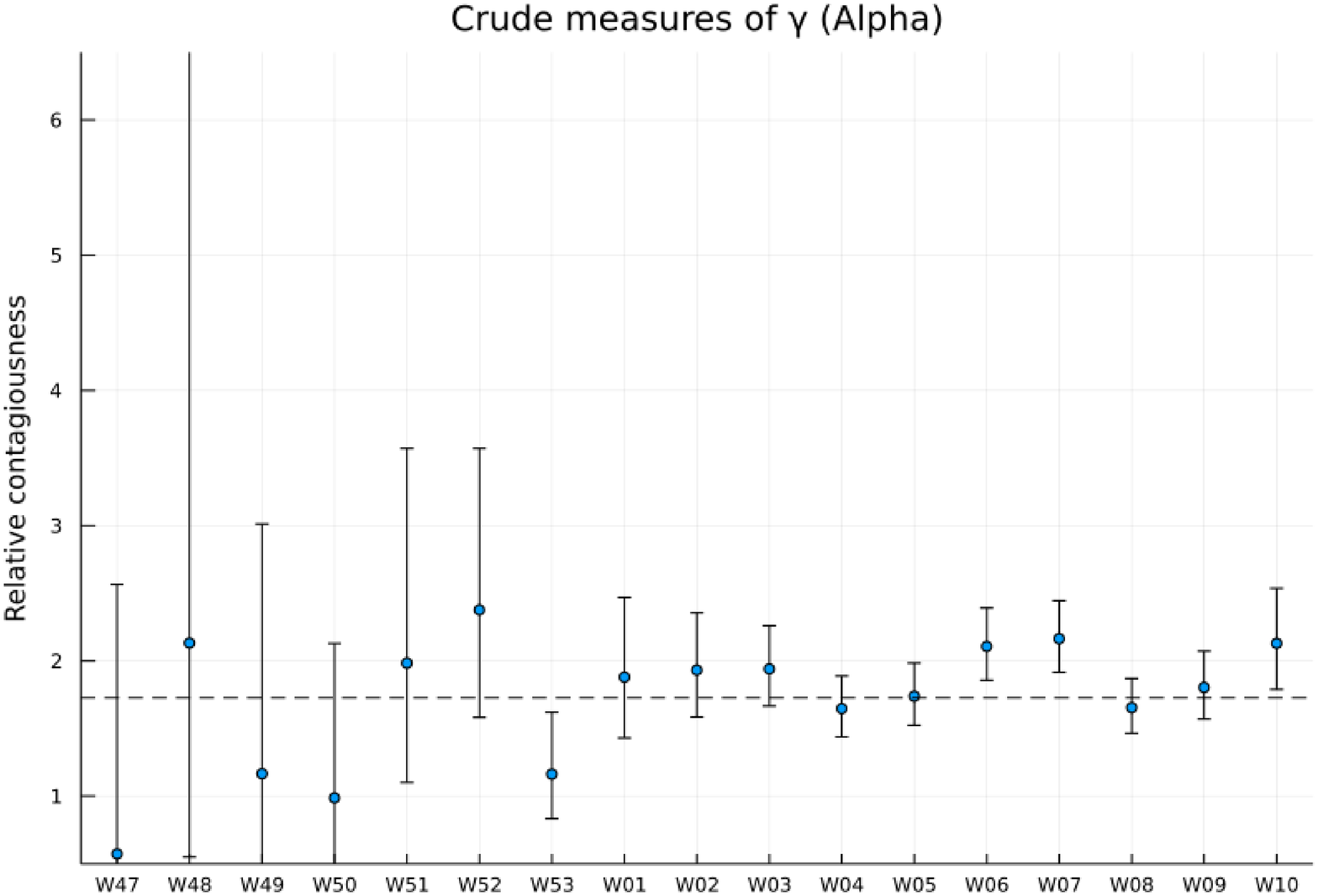}\includegraphics[viewport=0bp 5bp 450bp 542bp,height=0.3\textheight]{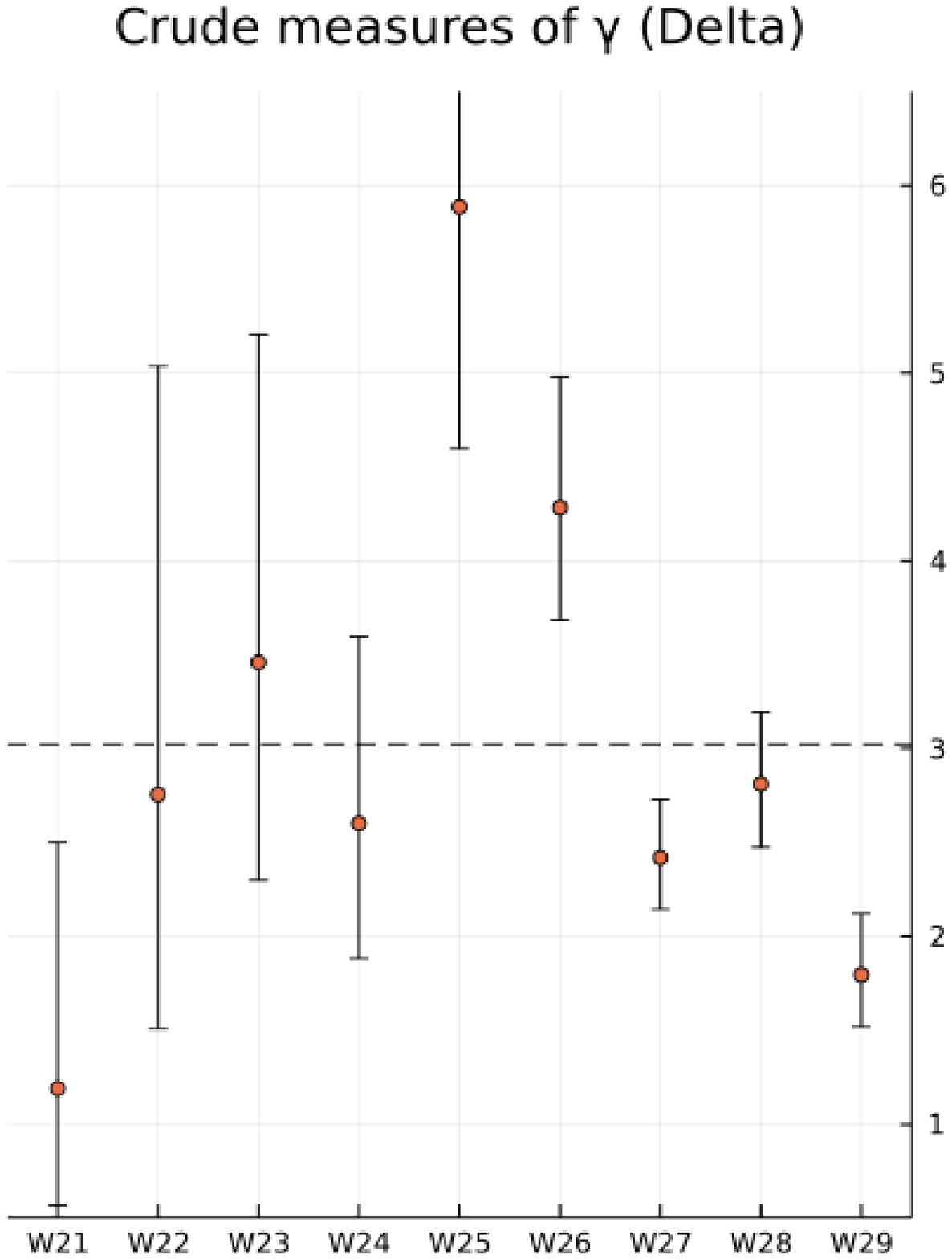}
\par\end{centering}
\caption{Crude weekly measures of the competitive advantage of the new variant.\label{fig:CrudeGamma}}
\end{figure}

A preliminary probing of the data can be done by considering the empirical
odds ratios of new-variant cases to old-variant cases. This ratio
should be approximately proportional to $B_{t}/A_{t}$, such that
the ratio of consecutive odds ratios,
\[
\left.\frac{X_{t}}{N_{t}-X_{t}}\right/\frac{X_{t-1}}{N_{t-1}-X_{t-1}}=\frac{X_{t}/X_{t-1}}{(N_{t}-X_{t})/(N_{t-1}-X_{t-1})}\approx b_{t}/a_{t}=\gamma.
\]
Thus we can use the ratio of consecutive odds ratios as a measurements
of $\gamma$ in week $t$. These empirical ratios and the corresponding
confidence intervals are shown in Figure \ref{fig:CrudeGamma}.\footnote{Weekly crude measures for NHS England STP areas were reported in \citet[figure 3]{Volz2020.12.30.20249034},
who used the median as an estimate of the reproductive advantages.} The
horizontal dashed lines in Figure  \ref{fig:CrudeGamma} represent the average value of the crude measures, which is 1.73
in for Alpha and 3.19 for Delta. The crude measures tend to have large confidence intervals early
in the sample because the number of new-variant cases is small. The
width of the confidence intervals are also influenced by the number
of tests that are being sequenced, $N_{t}$. For instance, in week
25, this number was relatively small for the simple reason that there
were few positive COVID-19 cases in Denmark that week -- just 1,315
positive cases of which 1,165 were successfully sequenced. The crude
measures for the Alpha variant in the left panel of Figure \ref{fig:CrudeGamma}
stabilizes about their average value, 1.73. For the Delta variant,
the crude measures are substantially larger and more disperse. The
progression of the Delta variant was particularly rapid in weeks 25
and 26.

The crude measurements of $\gamma$ in Figure \ref{fig:CrudeGamma}
do not fully exploit the information in the data, and the simple sample
averages (the dotted lines in Figure \ref{fig:CrudeGamma}) do not
account for heteroskedasticity and autocorrelation in the measurements
errors. To exploit the information in full, we turn to maximum likelihood
estimation using the parametrization of the logistic regression. We
compute robust standard errors using the Parzen kernel (with bandwidth
parameter $K=4$), see the Appendix. The results are not very sensitive
to the choice of bandwidth, but the robust standard errors are somewhat
larger than the non-robust standard errors, especially for the Delta
variant, see Table \ref{tab:Sensitivity-Robustness}.

The maximum likelihood estimates along with 95\% confidence intervals
are presented in Table \ref{tab:EmpiricalEstimates}. The Alpha variant
is estimated to be about 86\% more contagious per week than the preceding
variant, which we refer to as the \emph{ancestral variant.}\footnote{The ancestral variant represents a group of variants without a WHO
label, with the most prevalent variant before Alpha being B.1.177
also know as 20E (EU1).} The Delta variant, which emerged after then Alpha variant had become
completely dominant, is estimated to be 216\% more contagious than
the Alpha variant on a weekly basis. The reproduction number for SARS-CoV-2
is defined for a generation period (the typical time from a person
gets infected to the same person infects the next person). For SARS-CoV-2
this period is shorter than a week. Statens Serum Institut in
Denmark uses 4.7 days as the generation period, which we adopt in our calculations.
We can convert $\hat{\gamma}$ to a period of $x$ days using $\gamma_{x\mathrm{days}}=\exp(\tfrac{x}{7}\log\gamma_{\mathrm{week}})$,
and the estimates for $x=4.7$ days are presented in the last row
of Table \ref{tab:EmpiricalEstimates}. The estimates suggest that
the Alpha variant has a reproduction number that is about 1.5 times
larger than the ancestral variant. The Delta variant is estimated
to increase the reproduction number by an additional factor of 2.17,
which implies more than a threefold increase relative to the ancestral
variant. This is in line with other estimates, which include those
for the Alpha variant based on British data by \citet{Volz2020.12.30.20249034}
and those for the Delta variant by \citet{Wenseleers:2021}. The implication
is that it requires a larger proportion ($1-1/R_{0}$) to be immune
to reach \emph{herd immunity}. Suppose that 70\% immunity was needed
for the ancestral variant. Our estimates of $\gamma_{4.7\mathrm{days}}$
suggest this number increased to about 80\% for the Alpha variant
and about 90\% for the Delta variant.
\begin{table}
\caption{Empirical estimates for Alpha and Delta.\label{tab:EmpiricalEstimates}}
\begin{center}
\begin{tabular*}{\textwidth}{cccccc}
\hline\hline 
\noalign{\vskip2mm}
 &  &  & Alpha vs Ancestral & $\quad$Delta vs Alpha$\quad$ & Delta vs Ancestral\tabularnewline[2mm]
\hline
\noalign{\vskip2mm}
\multicolumn{4}{l}{\emph{Per Week}} &  & \tabularnewline
\noalign{\vskip2mm}
 & $\alpha$ & $\qquad$  & $\underset{[-9.00,-8.50]}{-7.8}$ & $\underset{[-8.75,-6.87]}{-7.81}$ & \tabularnewline[2mm]
\noalign{\vskip2mm}
 & $\beta$ &  & $\underset{[0.601,0.636]}{0.619}$ & $\underset{[1.026,1.278]}{1.152}$ & \tabularnewline[2mm]
\noalign{\vskip2mm}
 & $\gamma_{\mathrm{week}}$ &  & $\underset{[1.82,1.89]}{1.86}$ & $\underset{[2.79,3.59]}{3.16}$ & $\underset{[5.17,6.67]}{5.87}$\tabularnewline[2mm]
\noalign{\vskip2mm}
\hline
\noalign{\vskip2mm}
\multicolumn{4}{l}{\emph{Per Generation (4.7 days)}} &  & \tabularnewline[2mm]
\noalign{\vskip2mm}
 & $\gamma_{4.7\mathrm{days}}$  &  & $\underset{[1.50,1.53]}{1.51}$ & $\underset{[1.99,2.36]}{2.17}$ & $\underset{[3.01,3.58]}{3.28}$\tabularnewline[2mm]
\noalign{\vskip2mm}
\hline\hline 
\end{tabular*}
\end{center}
\footnotesize
     \renewcommand{\baselineskip}{11pt}
     \textbf{Note:} Empirical estimates with 95\% confidence intervals computed with robust
        standard errors. The estimates of relative contagiousness are $\gamma_{\mathrm{week}}=\exp(\beta)$ and
        $\gamma_{4.7\mathrm{days}}=\exp(\tfrac{4.7}{7}\beta)$.
\end{table}

\begin{figure}
\centering
\includegraphics[height=0.3\textheight]{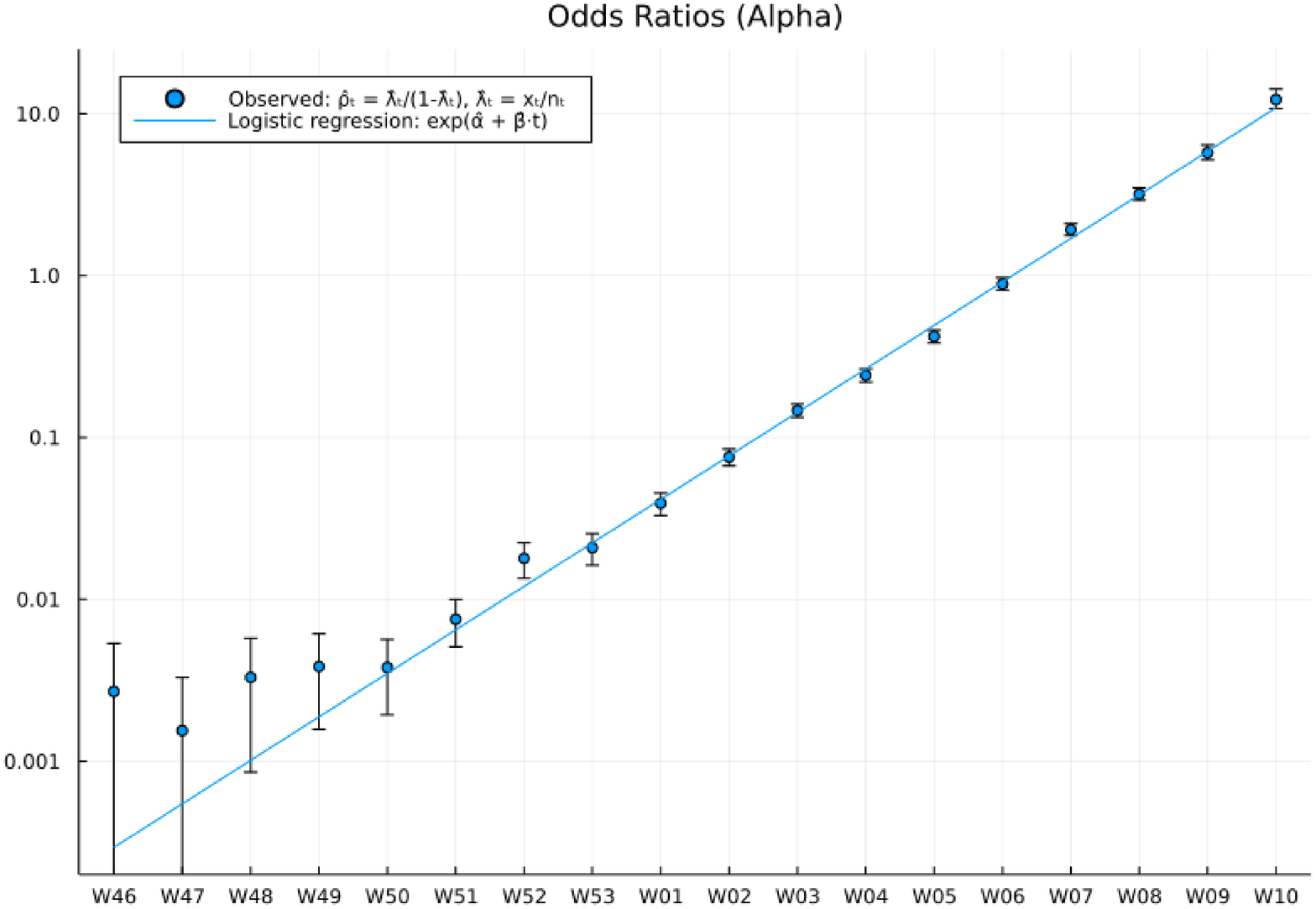}\includegraphics[viewport=0bp 4bp 450bp 544bp,height=0.3\textheight]{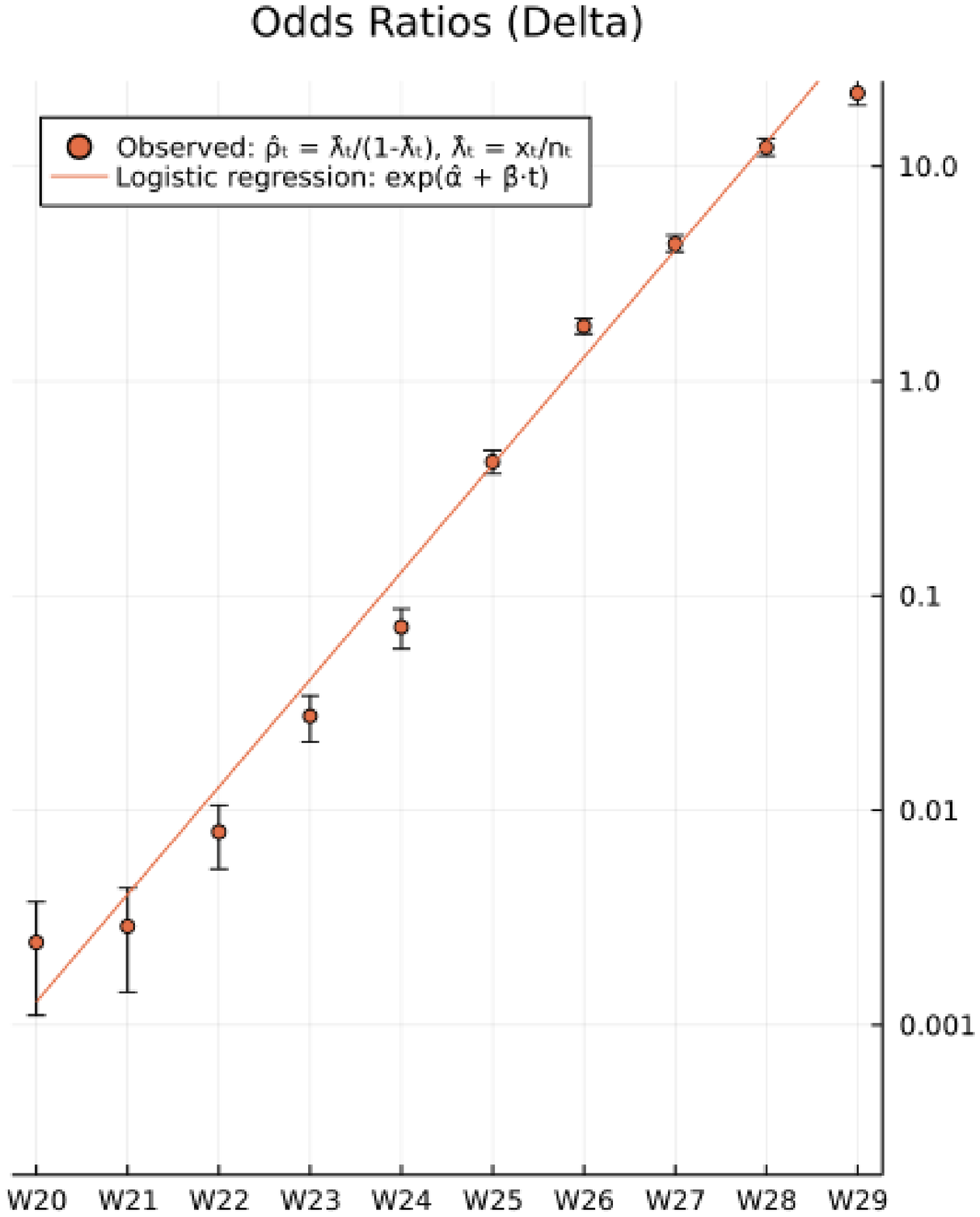}
\caption{Observed and estimated odds ratios.\label{fig:LogOddsRatios}}
\end{figure}

The estimated model and the observed odds ratios are shown in Figure
\ref{fig:LogOddsRatios}. Overall the model fit looks good, especially
for the analysis of the Alpha variant. There are some discrepancies
between the data and the linear specification for log odds ratios
with the Delta variant. A possible explanation is that many of the
COVID-19 cases that were detected in Denmark during the second sample
period were contracted abroad. According to the Danish Patient Safety
Authority, about 25\% of Covid-19 cases were imported cases, primarily
by people who had been vacationing in Spain in July.\footnote{https://www.ssi.dk/aktuelt/nyheder/2021/en-stor-del-af-covid-19-smitten-i-danmark-kommer-fra-de-rejsende}
This could potentially influence the progression of the Delta variant
because imported cases could be acquired in areas with a higher or
a lower Delta proportion than that in Denmark.

A second possible explanation is that the relative contagiousness
was different for unvaccinated and vaccinated individuals, causing
a gradual shift in the progression of the Delta variant as the vaccine
coverage increased. A study based on breakthrough infections in Denmark
during the period from March 1, 2021 to August 3, 2021 found a small
decline in vaccine effectiveness against the Delta variant. The Pfizer
vaccine was reported to be 81.0\% (95\% CI: 79.4; 82.4) effective
at preventing Alpha variant infections and 78.8\% (95\% CI: 77.2;
80.4) effective at preventing Delta variant infections.\footnote{https://www.ssi.dk/aktuelt/nyheder/2021/der-er-fortsat-hoj-vaccineeffektivitet-for-covid-19-vaccinerne}
The Pfizer vaccine (Comirnaty) is the most commonly used vaccine in
Denmark and accounts for about 85\% of all vaccinations. The study
only reported vaccine efficacy for fully vaccinated individuals, which
only accounted for a small percentage of individuals during this period.
A large discrepancy in efficacy between the Alpha and Delta variants
following a single vaccine dose could potentially have influenced
our analysis. However, the right panel of Figure \ref{fig:LogOddsRatios}
does not suggest a gradual change in the slope but a minor parallel
shift around Week 25.

A third possible explanation is that restrictions were largely abolished
during the second sample period, which could result in more noisy
data for the Delta variant and possible misspecification of the model.
During the Alpha sample period restrictions were quite restrictive.
In contrast, during the Delta sample period most restrictions were
abolished in Denmark, especially in relation to large gatherings.
The relaxed restrictions may explain the larger degree of randomness
in the progression of the Delta variant. For instance, the Euro 2020
games in Copenhagen may have contributed to the accelerated growth
in the Delta variant in Week 25 (see right panel of Figure \ref{fig:LogOddsRatios})
because spectators at two games accounted for a large fraction of
the Delta variant cases. Following the Denmark-Belgium Euro 2020 game
in Copenhagen on June 17, 2021, 41 attending spectators tested positive
for COVID-19 of which 25 cases (61.0\%) were the Delta variant. The
following week, on Monday June 21, 2021, Denmark played Russia in
Copenhagen at another Euro 2020 game, where 62 cases were subsequently
detected among spectators of which 28 (45.2\%) were Delta variant
cases. These are large numbers and percentages, because the total
number of Delta variant cases in Week 24 and Week 25 were 91 and 345,
respectively, and Delta variant only accounted for 6.7\% in Week 24
and 29.6\% in Week 25. The binomial model for Delta variant cases,
assumes that the individual cases are generated by independent Bernoulli
random variables. This independence assumption becomes questionable
when a large proportion of cases can be linked to the same events.
The right panel of Figure \ref{fig:LogOddsRatios} is consistent with
a single larger than expected jump in the proportion of Delta cases
around the time of the Euro 2020 games, and two parallel lines (one
fitting data up until Week 24 and one fitting data from Week 25 and
onwards) would appear to fit the data about as well as a single line
can fits the data in the left panel of Figure \ref{fig:LogOddsRatios}.
\begin{table}[!ht]
\caption{Daily COVID-19 PCR tests and outcomes (Omicron)} 
\begin{center}
\begin{tabular*}{\textwidth}{c@{\extracolsep{25pt}}c@{\extracolsep{\fill}}c@{\extracolsep{\fill}}r@{\extracolsep{0pt}}c@{\extracolsep{-2pt}}c@{\extracolsep{0pt}}c}
\hline\hline\\[-8pt] 
Date & Tested & Cases & \multicolumn{2}{c}{\begin{tabular}{c}Sequenced\\ (successfully)\end{tabular}} & \begin{tabular}{c}Omicron\\ cases\end{tabular} &  \begin{tabular}{c}Omicron\\ proportion\end{tabular}\\
 & (PCR) & $C_{t}$ &\multicolumn{1}{c}{$N_{t}$} &($N_{t}/C_{t}$)& $X_{t}$ & $X_{t}/N_{t}$\\[4pt]
\hline 
\hline 
2021-12-01  &   185,372 &   4,910   &   4,267  & (86.9\%) & 77  &    1.8\% \\
2021-12-02  &   213,494 &   5,040   &   4,294  & (85.2\%) & 62  &    1.4\% \\
2021-12-03  &   188,041 &   5,651   &   4,946  & (87.5\%) & 75  &    1.5\% \\
2021-12-04  &   140,790 &   5,577   &   5,089  & (91.2\%) & 111 &    2.2\% \\
2021-12-05  &   147,722 &   5,450   &   4,995  & (91.7\%) & 167 &    3.3\% \\
2021-12-06  &   209,434 &   7,645   &   6,762  & (88.4\%) & 337 &    5.0\% \\
2021-12-07  &   207,987 &   7,902   &   6,928  & (87.7\%) & 515 &    7.4\% \\
2021-12-08  &   205,263 &   7,136   &   6,232  & (87.3\%) & 649 &   10.4\% \\
2021-12-09  &   243,089 &   7,157   &   6,228  & (87.0\%) & 707 &   11.4\% \\
2021-12-10  &   210,756 &   7,520   &   6,444  & (85.7\%) & 843 &   13.1\% \\
2021-12-11  &   153,995 &   7,210   &   6,443  & (89.4\%) & 1,080   &   16.8\% \\
2021-12-12  &   165,474 &   7,723   &   6,794  & (88.0\%) & 1,521   &   22.4\% \\
2021-12-13  &   229,948 &  11,350   &   9,316  & (82.1\%)   & 2,691 &   28.9\% \\
2021-12-14  &   221,944 &  12,252   &   10,456 & (85.3\%)   & 4,044 &   38.7\% \\
2021-12-15  &   217,007 &  12,041   &   10,409 & (86.4\%)   & 4,827 &   46.4\% \\
2021-12-16  &   254,680 &  11,388   &   9,475  & (83.2\%)   & 4,438 &   46.8\% \\
2021-12-17  &   233,617 &  11,950   &   9,860  & (82.5\%)   & 5,213 &   52.9\% \\
2021-12-18  &   174,168 &  11,420   &   9,233  & (80.8\%)   & 5,163 &   55.9\% \\
2021-12-19  &   180,302 &  11,717   &   7,927  & (67.7\%)   & 4,908 &   61.9\% \\
2021-12-20  &   267,264 &  15,228   &   2,565  & (16.8\%)   & 1,611 &   62.8\% \\
2021-12-21  &   254,893 &  14,875   &   3,199  & (21.5\%)   & 2,437 &   76.2\% \\
2021-12-22  &   269,139 &  13,684   &   1,323  &  (9.7\%)   & 1,035 &   78.2\% \\
2021-12-23  &   243,139 &  14,729   &   3,450  & (23.4\%)   & 2,708 &   78.5\% \\
2021-12-24  &    71,463 &   8,322   &   597    &   (7.2\%)  &   494 &   82.7\% \\
2021-12-25  &    71,502 &   9,233   &   915    &   (9.9\%)  &   705 &   77.0\% \\
2021-12-26  &    79,592 &  12,300   &   2,297  & (18.7\%)   & 1,986 &   86.5\% \\
2021-12-27  &   182,893 &  25,168   &   4,657  & (18.5\%)   & 4,134 &   88.8\% \\
2021-12-28  &   191,226 &  24,273   &   1,471  &  (6.1\%)   & 1,324 &   90.0\% \\
2021-12-29  &   213,584 &  19,292   &   359    &   (1.9\%)  &   333 &   92.8\% \\
2021-12-30  &   225,529 &  21,727   &   910    &   (4.2\%)  &   829 &   91.1\% \\
2021-12-31  &    71,125 &  11,027   &   429    &   (3.9\%)  &   393 &   91.6\% \\
\hline\hline 
\end{tabular*}
\end{center}
\footnotesize
     \renewcommand{\baselineskip}{11pt}
     \textbf{Note:} Source is \emph{Variant-PCR svar fra 27. nov. og frem, Testcenter Danmark} SSI, January 16, 2022. Data available at: https://files.ssi.dk/covid19/podepind-sekventering/variant-pcr-test-december2021/opgoerelse-variantpcr-covid19-16012022-q9zx\label{tab:Daily-Danish-data}
\end{table}

\subsection{Omicron Variant}

A earlier version of this paper, dated November 7, 2021, analyzed the Alpha and Delta variants and concluded: ``It is unclear when a more contagious variant will emerge, if at all". This was resolved unequivocally a few weeks later with the arrival of the Omicron variant.
Below we analyze the competitive advantage of the Omicron variant and we will use daily data because this variant progressed to become dominant even faster than earlier variants.

Several notable changes occurred during the period where Omicron emerged. With the emergence of Omicron along with a rapid increase in COVID-19 case numbers, the Danish health authorities rapidly expanded the booster vaccination program (3rd dose of an mRNA vaccine) and recommended several preventive measures to slow the spread of the virus. In December alone, 1,960,421 individuals received a COVID-19 booster vaccine, which was in addition to 847,187 individuals who had received the booster vaccine before December 1, 2021.\footnote{Source: SSI. https://covid19.ssi.dk/overvagningsdata/download-fil-med-vaccinationsdata} Effective December 10th, mask requirements were reintroduced along with new restrictions on nightlife, restaurants, and indoor gatherings. Moreover, school children went back to remote learning on December 15th. 

The daily data for December 2021 are given in Table \ref{tab:Daily-Danish-data}. Nearly 6 million COVID-19 PCR tests\footnote{An additional 6.8 million antigen tests were performed during the same months. A positive antigen test is usually supplemented with a PCR test, from which the variant can be determined.} were performed during the month of December and the variant was determined in 158,270 of 350,897 positive test. Initially SSI attempted to sequence nearly all positive tests, but was unable to keep up with many cases and began sequencing a representative sample in late December.\footnote{It is unclear how the representative sampling is conducted and whether it is properly stratified by region, age, and vaccination status. SSI states: "The proportion of Omicron cases is calculated using only omicron variant PCR tested
samples from Test Center Denmark. We expect that the calculated proportion is
representative for all the confirmed SARS-CoV-2 cases."}  The Omicron proportion proportion grew from 10.4\% on December 8th to 90\% on December 28th and COVID-19 cases more than tripled in this period.

The maximum likelihood estimates along with 95\% confidence intervals are presented in Table \ref{tab:EmpiricalEstimatesOmicron}.\footnote{The robust standard errors are computed with the Parzen kernel and bandwidth parameter $K=4$, as was the case in the analysis of the Alpha and Delta variants.} Just prior to the emergence of Omicron, the Delta variant accounted for virtually all COVID-19 cases in Denmark. The estimates can therefore be interpreted as the competitive advantage of Omicron relative to Delta in a population where about 80\% are fully vaccinated. The latter has important implications for the external validity of these empirical results. The competitive advantages estimated for Alpha and Delta can be translated to larger basic reproduction numbers, because these variants emerged in a population with little prior immunity. In contrast, Omicron emerged in populations with a high degree of prior immunity from vaccinations and prior infections, and Omicron's ability to evade prior immunity explains a large part its competitive advantage over Delta. It is therefore not clear (from these data alone) if the basic reproduction number is substantially larger for Omicron than for Delta. The competitive advantage of Omicron could therefore vary across population with different degrees of prior immunity and between regions with different restrictions. In the Danish population with high vaccine coverage, Omicron is estimated to be 28\% more competitive than Delta \emph{per day}, which translate to 5.5 times more competitive per week. For the sake of comparison, we also report the competitive advantage of the Omicron variant \emph{per 4.7 days}, although the generation time for Omicron may be shorter than that of Delta. This estimate is in line with \citet{NishiuraEtAl2022}, who also noted that the increase in the basic reproduction number (for Omicron relative to Delta) ``is likely very small, e.g. in the order in the order of 10-20\%".
\begin{table}[!ht]
\caption{Empirical estimates for Omicron (vs Delta).\label{tab:EmpiricalEstimatesOmicron}}
\begin{center}
\begin{tabular*}{\textwidth}{l@{\extracolsep{\fill}}c@{\extracolsep{\fill}}c@{\extracolsep{\fill}}c@{\extracolsep{8pt}}c}
\hline\hline 
\noalign{\vskip2mm}
&  $\alpha$ & $\beta$ & $\gamma$ \\[2mm]
\noalign{\vskip2mm}
{\emph{Per Day}}   & $\underset{[-4.49,-3.74]}{-4.11}$  & $\underset{[0.221,0.267]}{0.244}$  & $\underset{[1.25,1.31]}{1.28}$& \\[2mm]
\noalign{\vskip2mm}
\multicolumn{1}{l}{\emph{Per Generation (4.7days)}}     & &   & $\underset{[2.83,3.50]}{3.15}$& \\[2mm]
\noalign{\vskip2mm}
\multicolumn{1}{l}{\emph{Per Week}}   &  &   & $\underset{[4.70,6.47]}{5.52}$& \\[2mm]
\noalign{\vskip2mm}
\hline\hline 
\end{tabular*}
\end{center}
\footnotesize
     \renewcommand{\baselineskip}{11pt}
     \textbf{Note:} Empirical estimates with 95\% confidence intervals computed with robust
        standard errors.
\end{table}

\begin{figure}
\centering
\subfloat[Daily proportion of Omicron]{
\includegraphics[width=0.6\textwidth]{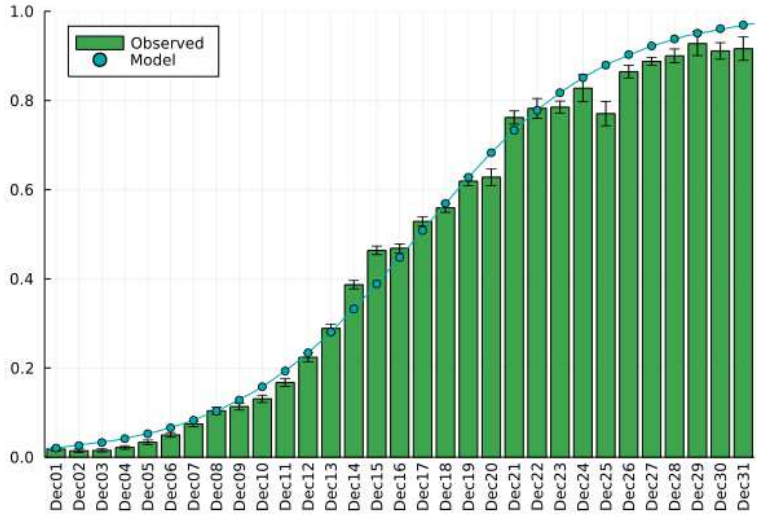}}
\subfloat[Daily Progression of Omicron]{
\includegraphics[width=0.4\textwidth]{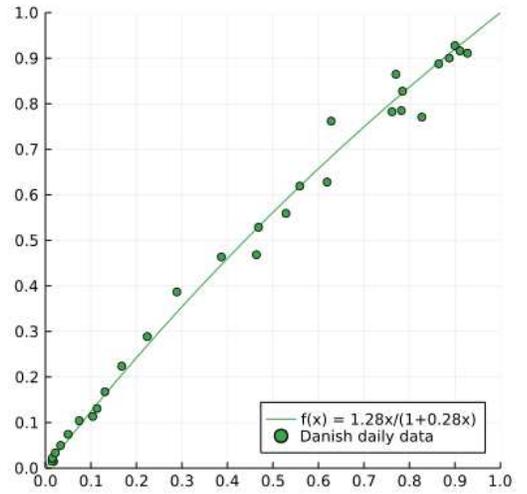}}
\par
\subfloat[Crude Measures ]{
\includegraphics[width=0.6\textwidth]{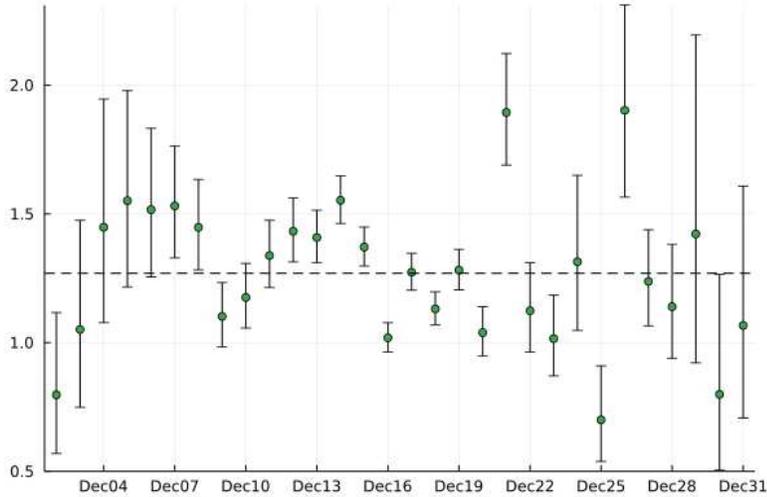}}
\subfloat[Daily Odds Ratios]{
\includegraphics[width=0.4\textwidth]{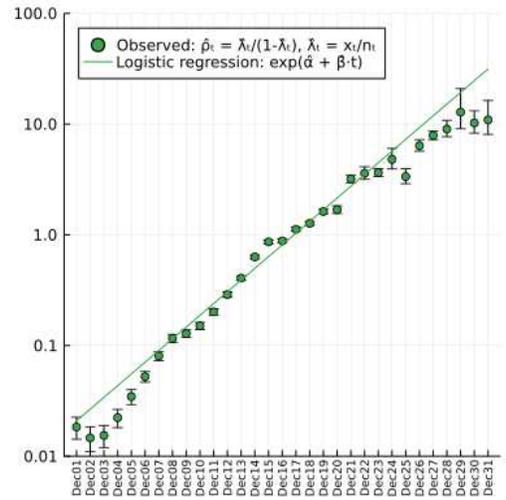}}
\par
\caption{Figures with empirical results for the Omicron variant.\label{fig:Omicron}}
\end{figure}

The four panels of Figure \ref{fig:Omicron} present results for Omicron that are analogous to the results presented for Alpha and Delta in Figures 1-4. Panel (a) displays the daily proportion of Omicron in positive PCR tests that were successfully sequenced and the solid line is the proportion implied by the estimated model. The model tracks the observed proportions reasonably well with the exception of the last week where it overpredicts the proportion of Omicron. Below, we discuss possible explanations for this and other discrepancies. Panel (b) in Figure \ref{fig:Omicron} shows the daily progression implied by a daily competitive advantage of $\gamma=1.28$. The crude measures in Panel (c) are more erratic than the crude measures based on the weekly data in Figure \ref{fig:CrudeGamma}. The dashed line represents their average value, which (at 1.27) is close to the maximum likelihood estimate.  Panel (d) displays the estimated model for the daily odds ratios, $\rho_t$, and the observed odds ratios. The estimate of $\gamma$ defines the slope of the fitted line in Panel (d). The observed data suggests some time variation in $\gamma$, including weekly periodicity, which can also be seen in Panel (c). Many of the Omicron cases in early December were traced to super spreader events, which had occurred during weekend activities and it is interesting to note that the odds ratios for Omicron tend to be relatively large on Wednesdays (December 1, 8, 15, 22, and 29), because Wednesdays are days where individuals infected during the weekend are likely to test positive. The periodic variation across day-of the-week may be explained by Omicron having a particular competitive advantage during events that tend to occur during weekends. Another possible explanation is that Omicron was more prevalent among young adults, and these age cohorts accounted for a disproportional large share of infections that were detected on Wednesdays. More detailed data where case numbers are stratified by both variant and age could cast light on this aspect.

The progression of Omicron is below the estimated trendline towards the end of December. The possible explanations for this are many. First, the Christmas holiday may have played a role, for instance by changing contact patterns and increased use of preventive testing before seeing relatives. Second, it is also possible that the sampling of positive PCR tests, which was adopted in late December, did not succeed in being representative. Third, it is also possible that restrictions introduced in mid December were relatively more effective at preventing Omicron transmissions than Delta transmissions. A fourth possible explanation is that booster vaccinations induced time variation in $\gamma$. There was a large shift in population immunity during the month of December, as a large share of the population received a booster vaccine. The competitive advantage of Omicron might be smaller in this subpopulation and this is supported by the evidence that boosters vaccination greatly reduce the risk of Omicron infections, see \citet{LyngseEtAl2022}, and the observation that the percentage of Omicron cases increased in the unvaccinated group during the month of December. It would be interesting to generalized the model framework to encompass subpopulations with different values of $\gamma$. This would require a careful modeling of the interactions between groups and data with the required granularity.

\section{Confidence Intervals, Predictions, and Inferring reproduction Number}

In this section, we detail detail two ancillary results. First, in
Section 4.1, we show the estimated model can be used to predict the
proportion of an emerging virus variant and develop methods for quantifying
the associated uncertainty. We illustrate these methods with the data
for the Alpha variant. Then, in Section 4.2, we develop a simple formula
for the reproduction number of the new variant, which does not require
concurrent genome data. Instead it projects the most recent estimate
of the proportion forward and infer the effective reproduction number
from the recent growth in total cases.

\subsection{Confidence Sets and Out-of-Sample Analysis}

At times $T$ we can estimate $\alpha$ and $\beta$, as well as their
variance-covariance matrix, $\Sigma_{T}=\mathrm{var}((\hat{\alpha}_{T},\hat{\beta}_{T})^{\prime})$,
where $\hat{\alpha}_{T}$, $\hat{\beta}_{T}$, and $\hat{\Sigma}_{T}$
denote the resulting estimates. Point forecasts for the proportion
of the new virus variant, $\lambda_{t}$, are given from (\ref{eq:LogisticReg}).
The $h$ period ahead point forecast, made at time $T$, is simply
\[
\hat{\lambda}_{T+h,T}=1\left/\left[1+\exp\{-\hat{\alpha}_{T}-\hat{\beta}_{T}(T+h)\right]\right.,
\]
and the corresponding confidence bands can be deduced from the asymptotic
distribution of $(\hat{\alpha}_{T},\hat{\beta}_{T})$.
The confidence
band based on $c$ units of standard deviations is given by 
\begin{equation}
1\left/\left[1+\exp\{-\hat{\alpha}_{T}-\hat{\beta}_{T}(T+h)\pm c\sqrt{v(T+h,\hat{\Sigma}_{T})}\}\right]\right.,\label{eq:lambdaUncertainty}
\end{equation}
where $c=1.96$ would correspond to a 95\% confidence bands and
\begin{figure}
\centering
\includegraphics[width=0.45\textwidth]{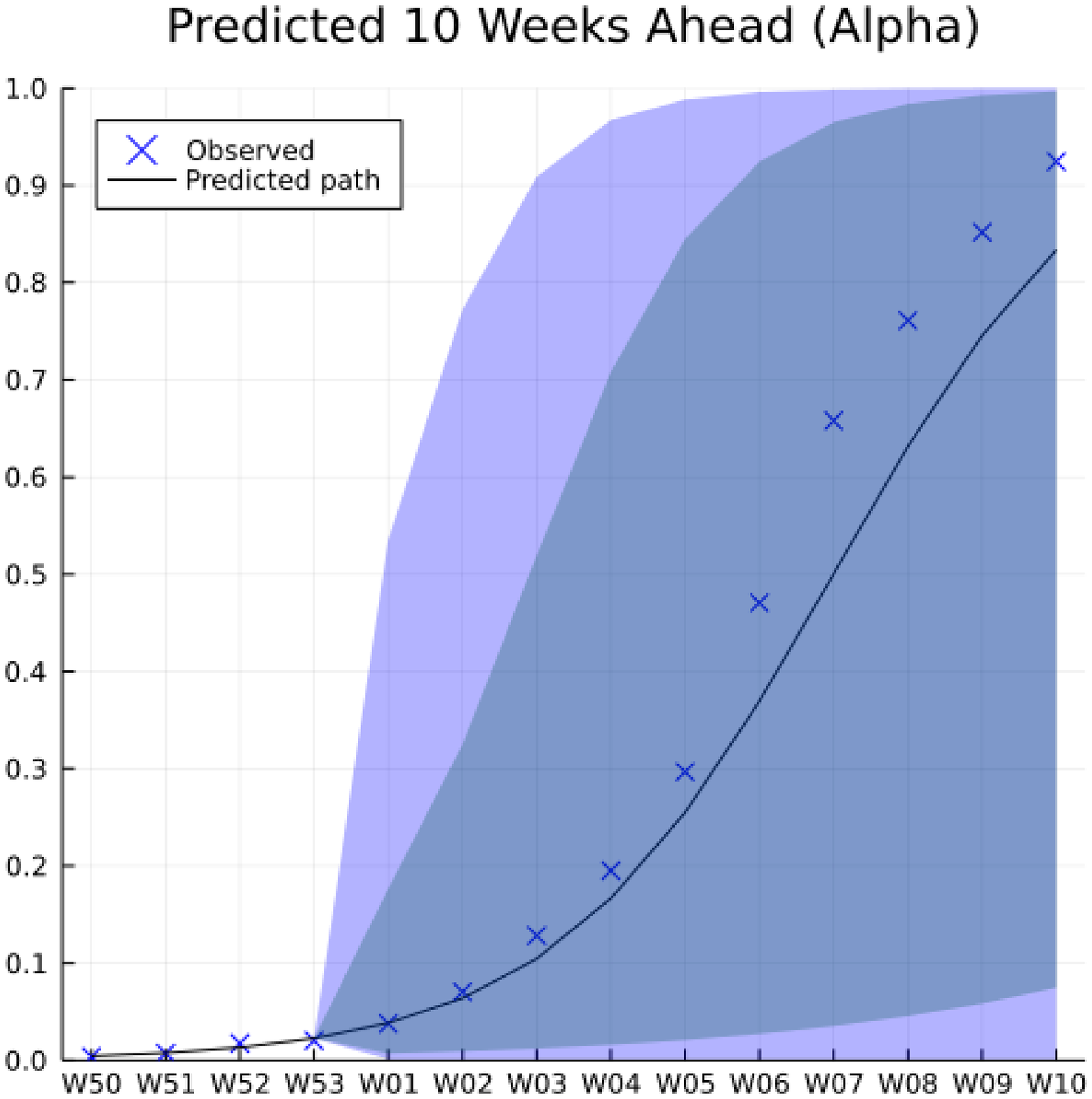}\includegraphics[width=0.45\textwidth]{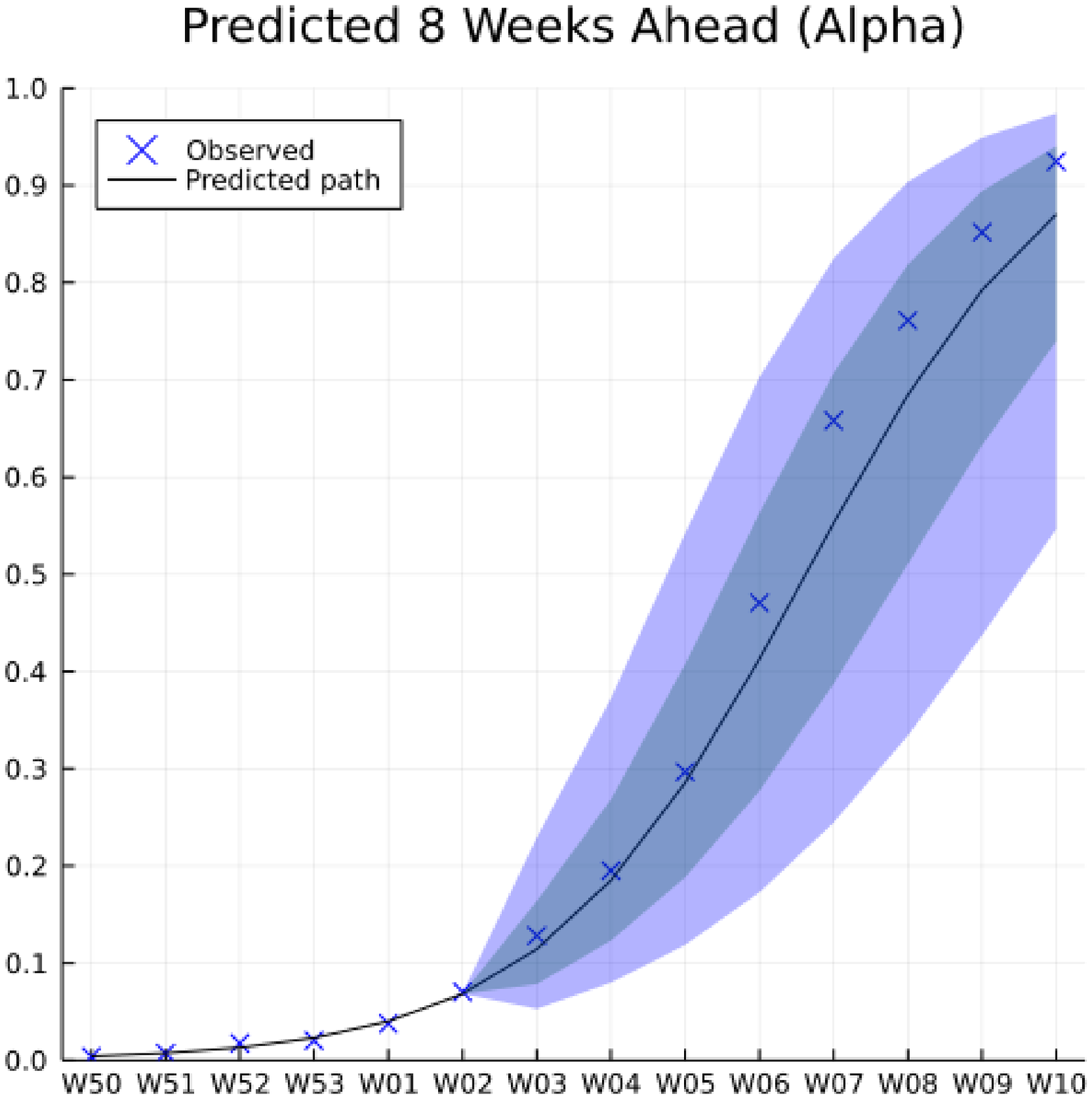}
\par
\includegraphics[width=0.45\textwidth]{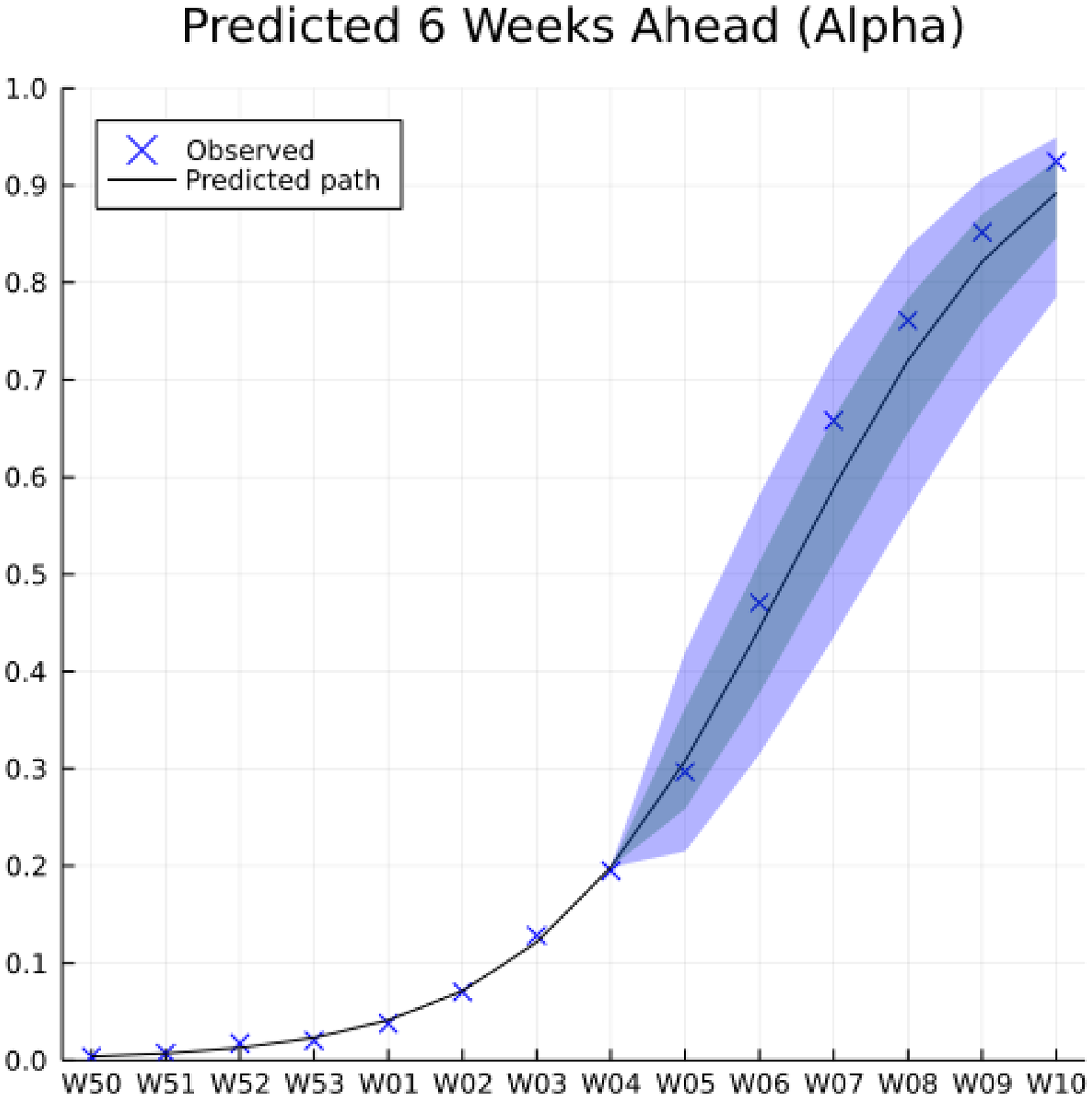}\includegraphics[width=0.45\textwidth]{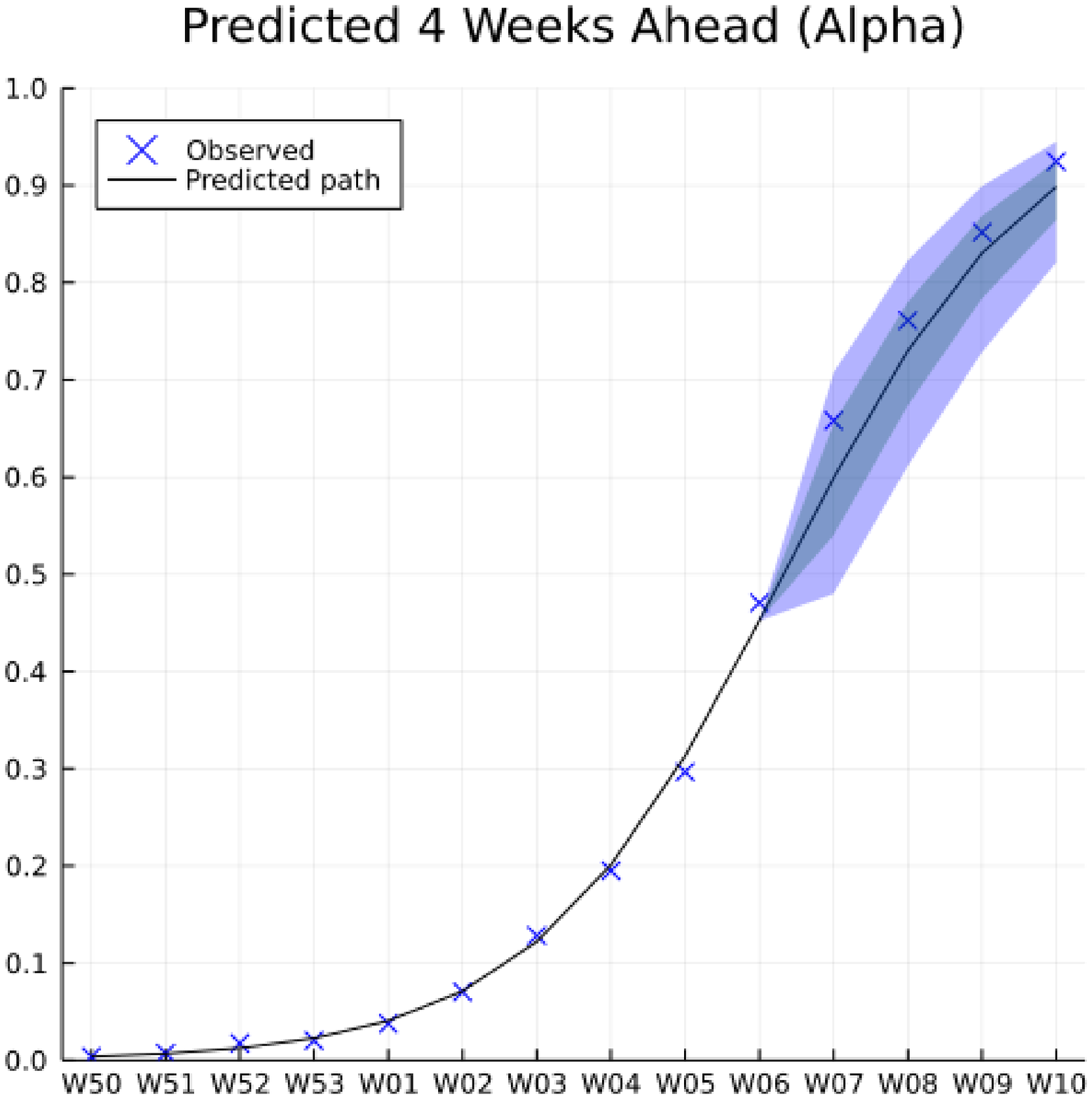}
\caption{The predicted path for $\lambda_{t}$ (solid black line) is shown
for when the model is estimated with, 4, 6, 8, and 10 weeks of data,
which translates to an out-of-sample period of 10, 8, 6 and 4 weeks,
respectively. The shaded areas are the confidence bands using $2\times$
and $4\times$ the standard deviation as defined in (\ref{eq:lambdaUncertainty}).
The observed proportion of Alpha are indicated with the blue crosses.\label{fig:lambda-function}}
\end{figure}
\[
v(T+h,\hat{\Sigma}_{T})\equiv\hat{\mathrm{var}}(\hat{\alpha}_{T}+\hat{\beta}_{T}(T+h))=(1,T+h)\hat{\Sigma}_{T}(1,T+h)^{\prime}.
\]
The estimated and predicted progression of $\lambda_{t}$ for the
Alpha variant along with confidence bands (using $c=2$ and $c=4$
standard deviations) are presented in Figure \ref{fig:lambda-function}.
The saltires (x-crosses) in Figure \ref{fig:lambda-function} are
the observed weekly empirical proportion of the Alpha variant.

In the upper left panel of Figure \ref{fig:lambda-function}, we have
estimated the model by maximum likelihood using 4 weeks of data (Week
50-53) which leaves 10 weeks for out-of-sample forecasting. The point
forecasts are reasonably close to the realized proportions, but with
just four weeks of data for estimation, there is a great deal of uncertainty
about the estimated parameters, causing $v(T+h,\hat{\Sigma}_{T})$
to be large. With two additional weeks for estimation (six weeks total),
the parameters are more precisely estimated, resulting in tighter
confidence bands, as shown in the upper-right panel of Figure \ref{fig:lambda-function}.
With eight or ten weeks for estimation, the parameter estimates become
even more accurate, resulting in the even tighter confidence intervals
in the two lower panels.
The point forecasts are reasonably accurate at horizons up to four
weeks, but tend to be below the realized values, especially at longer
horizons. This is because the four in-sample estimates of $\gamma$
($\hat{\gamma}_{W53}=1.71$, $\hat{\gamma}_{W02}=1.76$, $\hat{\gamma}_{W04}=1.79$,
and $\hat{\gamma}_{W06}=1.81$) are all smaller than the full sample
estimate: $\hat{\gamma}=1.86$. This highlights that we should expect
the out-of-sample forecasting errors to be positively autocorrelated
and likely have the same sign as $\gamma-\hat{\gamma}_{T}$. It should
be noted that the confidence bands reflect the uncertainty about $\lambda_{T+h}$,
while the realized empirical proportions, $X_{T+h}/N_{T+h}$, are
themselves noisy estimates of $\lambda_{T+h}$, see the confidence
bands in Figure \ref{fig:ProportionAlphaAndDelta}.

\subsection{Inferring the Reproduction Number for the New Variant in Real Time}

We can infer the effective reproduction number for an emerging variant
from the effective reproduction number of all cases when combined
with knowledge about $\lambda$ and $\gamma$. Let $C$ be the number
of all cases in this period, of which $B=\lambda C$ are the new-variant
cases and $A=(1-\lambda)C$ are the old-variant cases. If the current
reproduction number for all cases is $R$, then there were $C/R$
cases one generation ago. Similarly, there were $B/R_{B}=\lambda C/R_{B}$
new-variant cases and $A/R_{A}=(1-\lambda)C/R_{A}$ old-variant cases
one generation earlier, where $R_{A}$ and $R_{B}$ denote the current
reproduction numbers for the old and new variant, respectively. The
number of cases for the previous generation have to add up to the
total number of cases. Hence, $C/R=\lambda C/R_{B}+(1-\lambda)C/R_{A}$,
and since $R_{A}=R_{B}/\gamma$ it follows that 
\begin{equation}
R_{B}=R_{B}(\lambda,R,\gamma)=R[\lambda+\gamma(1-\lambda)].\label{eq:Infer-Rb}
\end{equation}
 The value of $\lambda$ to be used in this expression should be the
that for the current period, which is typically predicted from earlier
periods, and the value of $\gamma$ to be used in (\ref{eq:Infer-Rb}),
should be the one that corresponds to the same generation period as
used to compute $R$. We estimated $\gamma_{4.7\mathrm{days}}\approx1.5$
for the Alpha variant and $\gamma_{4.7\mathrm{days}}\approx2$ for
Delta. Thus, based on the Danish data we approximately have, 
\[
R_{\mathrm{Alpha},T}\approx R_{T}\times(1.5-0.5\lambda_{T})\qquad\text{and}\qquad R_{\mathrm{Delta},T}\approx R_{T}\times(2-\lambda_{T}).
\]
This formula makes it possible to assess the reproduction number for
an emerging variant before concurrent sequencing data are available.
The reproduction number, $R_{T}$, for all cases can be inferred from
the progression in the total number of COVID-19 cases and the proportion
of the new variant, $\lambda_{T}$, can be obtained from the estimated
model, by projecting forward the most recent knowledge about the proportion,
see Figure \ref{fig:LogOddsRatios}.

\subsubsection{Empirical Illustration for the Alpha Variant}

\begin{figure}
\centering
\includegraphics[width=0.9\textwidth]{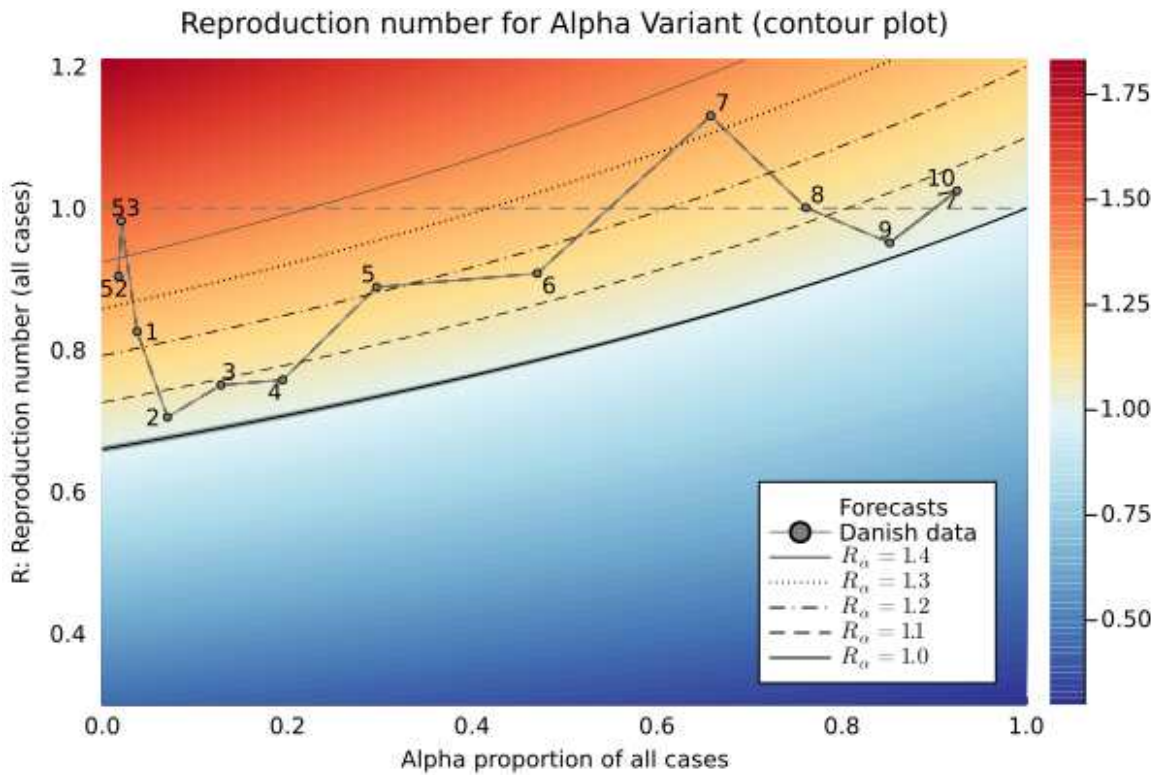}
\caption{Contour plot of weekly reproduction numbers for Alpha.\label{fig:StabilityRegion}}
\end{figure}

We can use (\ref{eq:Infer-Rb}) to characterize the combinations of
$(\lambda,R)$ that correspond to a particular reproduction number
for the Alpha variant. A contour plot for $R_{B}(\lambda,R)$ based
on the point estimate of $\gamma$ that corresponds to a generation
period, $\hat{\gamma}_{4.7\mathrm{days}}=\exp(\tfrac{4.7}{7}\log\hat{\gamma})$,
is presented in Figure \ref{fig:StabilityRegion}. The region above
the solid line, $\{(\lambda,c):R_{B}(\lambda,R,\hat{\gamma}_{4.7\mathrm{days}})>1\}$,
are the combinations of $\lambda$ and $R$ where case numbers for
Alpha are increasing, and the region below the solid line is the region
where Alpha cases are decreasing. The shaded region about the solid
line represent the uncertainty about the threshold, due to uncertainty
about $\gamma$. The shaded area is given by 
\[
\{(\lambda,R):R_{B}(\lambda,R,\gamma)=1,\text{ for some }\gamma\in\mathrm{CI}_{95\%}\},
\]
where $\mathrm{CI}_{95\%}$ is the 95\% confidence interval for $\hat{\gamma}_{4.7\mathrm{days}}$
we obtained in Section \ref{sec:Empirical-Analysis}. Note that the
uncertainty interval shrinks to zero as $\lambda\rightarrow1$. The
reason is that the limited case, $\lambda=1$, represents the situation
where Alpha cases make up all cases, and its rate of increase can
therefore be inferred from the rate of increase in all cases. More
formally, the result follows from the fact that $R_{B}/R=\lambda+\gamma(1-\lambda)\rightarrow1$
as $\lambda\rightarrow1$.

A model-free proxy for $\lambda_{t}$ is $X_{t}/N_{t}$ and a crude
estimate of $R$ in week $t$, is given by $\hat{R}_{t}\equiv\exp(\tfrac{4.7}{7}\log\frac{\mathrm{Cases}_{t}^{Adj}}{\mathrm{Cases}_{t-1}^{Adj}})$,
where $\mathrm{Cases}_{t}^{Adj}$ is the number of all cases in week
$t$ after adjusting for the testing intensity. The adjustment is
given by $\mathrm{Cases}_{t}^{Adj}=\mathrm{Cases}_{t}\times\left(\frac{\mathrm{Tested}_{t}}{M}\right)^{-0.7}$,
where $M$ is a baseline number of tests, see \citet{SSI_beta07_2020}.
The baseline number, $M$, does not influence the ratio because, 
\[
\frac{\mathrm{Cases}_{t}^{Adj}}{\mathrm{Cases}_{t-1}^{Adj}}=\frac{\mathrm{Cases}_{t}}{\mathrm{Cases}_{t-1}}\times\left(\frac{\mathrm{Tested}_{t}}{\mathrm{Tested}_{t-1}}\right)^{-0.7},
\]
and we use this ratio to compute $\hat{R}_{t}$.

The estimated reproduction number, $\hat{R}_{t}$, is plotted against
the observed proportion of the Alpha variant in Figure \ref{fig:StabilityRegion},
labelled with the corresponding week number. All pairs fall above
the solid line, where the effective reproduction number for the Alpha
variant is greater than one. This indicates that the number of Alpha
cases (detected and undetected) was growing throughout the sample
period even though the total number of cases was declining most weeks in this sample period.
 
\section{Discussion}

We have shown how the relative contagiousness of a new virus variant
can be estimated by maximum likelihood and how robust standard errors
can be computed. The underlying structure is that of a logistic regression
model. We applied the methodology to Danish data from the periods
where the Alpha, Delta, and Omicron variants emerge and become dominant.
The methodology can also be applied to data with irregular sampling frequencies,
such as time series with missing data, and the analysis
can be extended to situations with more than two competing
variants as detailed in Appendix A. The framework is not specific to the analysis of competing virus variant,
but can be applied in a context with other competing objects. We
found the Alpha variant increased the contagiousness by about 50\%
and the Delta variant increased the contagiousness further by more
than 100\% per generation. We estimated the Omicron variant to be about three 
times more infectious than the Delta variant in a highly vaccinated population.
The empirical results should be interpreted in relation to the data they were estimated with.
The results for Alpha and Delta were derived in a population that was largely immune naive. These findings have external validity, because they relates directly to the basic reproduction number of the virus. The results for Omicron do not have the same degree of external validity because the competitive advantage was estimated in a population with a particular (and changing) composition of immunity. 

Three new variants of the SARS-CoV-2 have emerged to become dominant
in short succession. There were just 18 weeks between
the time the Alpha variant made up 90\% of all cases in Denmark to the time the
Delta variant surpassed that same threshold, and 28 weeks later the Omicron variant accounted for over 90\% of all cases. 
It is therefore reasonable to expect that even more contagious variants
will emerge in the time to come. While the first two variants were not only more contagious
but also found to increase the risk of hospitalization. Fortunately, there is evidence that the Omicron variant is less severe and despite the many breakthrough infections there is strong evidence that vaccinations continue to abate transmissions and greatly reduce the risk of severe disease.

{\footnotesize{}\bibliographystyle{elsarticle-num-names}
\bibliography{references}
}{\footnotesize\par}

\appendix
\setcounter{equation}{0}\renewcommand{\theequation}{A.\arabic{equation}}
\setcounter{table}{0}\renewcommand{\thetable}{A.\arabic{table}}

\section*{Appendix A: The Case with Multiple Competing Virus Variants}

In this Appendix we consider the model extension to the case where
there may be more than two competing virus variant.
Although this was not relevant in the
analysis of the Danish date, other countries have reported a sizable proportion of three or more variants simultaneously.
For instance, France had four variants that each accounted for more than 5\% of all COVID-19 cases in the second half of June, 2021,
Delta (41\%), Alpha (29\%), Beta (19\%), and Gamma (7\%).\footnote{According to GISAID data reported on \tt{https://covariants.org/}}

For $j=1,\ldots,m$, we let $C_{j,t}$ and $\lambda_{j,t}=C_{j,t}/C_{t}$
denote the number of cases and the proportion of the $j$-th variant
at time $t$. Here $m$ is the number of virus variants and $C_{t}=\sum_{j=1}^{m}C_{j,t}$
is the total number of cases at time $t$. We measure the contagiousness
of all variants relative to the first variant ($j=1$) and let $a_{t+1}=C_{1,t+1}/C_{1,t}$
represent the progression in this variant. So, the first variant is
used as the numéraire and the relative contagiousness of other variants
is represented by the parameter, $\gamma_{j}$, $j=1,\ldots,m$ with
$\gamma_{1}=1$. Analogous to the case with two variants we now have                
\[
C_{j,t+1}=\gamma_{j}a_{t+1}C_{j,t},\qquad j=1,\ldots,m,
\]
and the evolution of the variant proportions are given by 
\begin{eqnarray}
\lambda_{j,t+1}=\frac{C_{j,t+1}}{C_{t+1}} & = & \frac{\gamma_{j}a_{t+1}C_{j,t}}{a_{t+1}C_{1,t}+\gamma_{2}a_{t+1}C_{2,t}+\cdots+\gamma_{m}a_{t+1}C_{m,t}}\nonumber \\
 & = & \frac{\gamma_{j}C_{j,t}/C_{t}}{C_{1,t}/C_{t}+\gamma_{2}C_{2,t}/C_{t}+\cdots+\gamma_{m}C_{m,t}/C_{t}}.\label{eq:lambda-multiple}\\
 & = & \frac{\gamma_{j}\lambda_{j,t}}{\lambda_{1,t}+\gamma_{2}\lambda_{2,t}+\cdots+\gamma_{m}\lambda_{m,t}},\qquad j=1,\ldots,m.\nonumber 
\end{eqnarray}

\subsection*{A.1 Estimation and Inference}

Let $X_{j,t}$ be the number of cases that are identified as the $j$-th
variant at times $t$, and assume that $(X_{1,t},\ldots,X_{m,t})$
is a representative sample of $(C_{1,t},\ldots,C_{m,t})$. The log-likelihood
is now given by,

\begin{equation}
\ell(\gamma,\lambda_{0})\propto\sum_{t=1}^{T}\sum_{j=1}^{m}X_{j,t}\log\lambda_{j,t},\label{eq:LogLmulti}
\end{equation}
where $\lambda_{t}$ evolves according to (\ref{eq:lambda-multiple})
and the two vectors of unknown parameters are $\gamma=(\gamma_{2},\ldots,\gamma_{m})^{\prime}$
and the initial value $\lambda_{0}\equiv(\lambda_{2,0},\ldots,\lambda_{m,0})^{\prime}$.
These can be estimated by maximum likelihood, $(\hat{\lambda},\hat{\gamma})=\arg\max_{\gamma,\lambda}\ell(\lambda,\gamma)$,
and confidence intervals for $\lambda$ and $\gamma$ can be obtained
with conventional methods. This model does not conform with the simple
logistic regression model we used in the case with two variant, but
could be cast as a multinomial logistic regression model.

\subsection*{A.2 Omitting Variants from the Analysis}

For some variants, the number of observed cases may be too small to
obtain reliable estimates of their relative contagiousness. An interesting
question is if omitting all but two variants from the analysis will
induce bias and/or inconsistencies in the estimated relative contagiousness.
This is fortunately not the case. Suppose we drop variants $j=3,\ldots,m$
from the analysis and proceed to estimate $\gamma_{2}$ solely from
data on the first two variant. Define $\tilde{\lambda}_{t}=C_{2,t}/(C_{1,t}+C_{2,t})$,
then from (\ref{eq:lambda-multiple}) it follows that 
\begin{eqnarray*}
\tilde{\lambda}_{t+1}\equiv\frac{\lambda_{2,t+1}}{\lambda_{1,t+1}+\lambda_{2,t+1}} & = & \frac{\frac{\gamma_{2}\lambda_{2,t}}{\lambda_{1,t}+\gamma_{2}\lambda_{2,t}+\cdots+\gamma_{m}\lambda_{m,t}}}{\frac{\lambda_{1,t}}{\lambda_{1,t}+\gamma_{2}\lambda_{2,t}+\cdots+\gamma_{m}\lambda_{m,t}}+\frac{\gamma_{2}\lambda_{2,t}}{\lambda_{1,t}+\gamma_{2}\lambda_{2,t}+\cdots+\gamma_{m}\lambda_{m,t}}}\\
 & = & \frac{\gamma_{2}\lambda_{2,t}}{\lambda_{1,t}+\gamma_{2}\lambda_{2,t}}=\frac{\gamma_{2}\frac{\lambda_{2,t}}{\lambda_{1,t}+\lambda_{2,t}}}{\frac{\lambda_{1,t}}{\lambda_{1,t}+\lambda_{2,t}}+\gamma_{2}\frac{\lambda_{2,t}}{\lambda_{1,t}+\lambda_{2,t}}}=\frac{\gamma_{2}\tilde{\lambda}_{t}}{(1-\tilde{\lambda}_{t})+\gamma_{2}\tilde{\lambda}_{t}},
\end{eqnarray*}
which is identical to (\ref{eq:lambda}). This shows that estimation
and inference about a single relative contagiousness parameter can
be based entirely on data forthe two variants whose relative contagiousness
is the object of interest. 

Forecasting, as well as most hypothesis tests involving multiple $\gamma$-parameters,
would require knowledge about the dependence between the estimates.
Such situations calls for estimation of the full model defined by
the log-likelihood in (\ref{eq:LogLmulti}). 

\section*{Appendix B: Robust Standard Errors of Estimators}

While the log-likelihood estimates are identical to those obtained
with logistic regression packages, the standard errors provided by
most packages are based on the Fisher Information matrix, $\hat{\mathcal{I}}$
(defined below), and for these to be reliable, the model must be correctly
specified. We will compute standard errors using the sandwich form
of variance-covariance matrix for the estimated parameters, which
is detailed next.

We parameterize the log-likelihood with the standard parameterization
of the logistic regression, $\theta^{\prime}=(\alpha,\beta)=(\log\rho_{0},\log\gamma)$.
The maximum likelihood estimates are obtained by maximizing $\sum_{t=1}^{T}\ell_{t}(\theta)$,
where $\ell_{t}(\theta)=X_{t}\log\lambda_{t}+(N_{t}-X_{t})\log(1-\lambda_{t})$.
To compute robust standard errors we derive the score , $s_{t}(\theta)=\frac{\partial\ell_{t}(\theta)}{\partial\theta}$,
and hessian, $h_{t}(\theta)=\frac{\partial^{2}\ell_{t}(\theta)}{\partial\theta\partial\theta^{\prime}}$.
To this end we observe that the derivatives of $\lambda_{t}(\theta)=\lambda_{t}(\alpha,\beta)=[1+e^{-\alpha-\beta t}]^{-1}$
are simply
\[
\frac{\partial\lambda_{t}(\theta)}{\partial\alpha}=-e^{-\alpha-\beta t}\frac{1}{(1+e^{-\alpha-\beta t})^{2}}=-\lambda_{t}(1-\lambda_{t}),
\]
and similarly of $\partial\lambda_{t}/\partial\beta=-\lambda_{t}(\theta)[1-\lambda_{t}(\theta)]\times t$,
such that the score for the observations in the $t$-th week is given
\begin{equation}
s_{t}(\theta)=[-X_{t}(1-\lambda_{t})+(N_{t}-X_{t})\lambda_{t}]\left[\begin{array}{c}
1\\
t
\end{array}\right]=(N_{t}\lambda_{t}-X_{t})\left[\begin{array}{c}
1\\
t
\end{array}\right].\label{eq:score}
\end{equation}
Next, by combining the expression for $\partial\lambda_{t}(\theta)/\partial\theta$
with (\ref{eq:score}) we obtain,
\[
h_{t}(\theta)=-N_{t}\lambda_{t}(1-\lambda_{t})\left[\begin{array}{cc}
1 & t\\
t & t^{2}
\end{array}\right].
\]
It is now straightforward to compute the information matrices $\hat{\mathcal{J}}=\sum_{t=1}^{T}s_{t}(\hat{\theta})s_{t}(\hat{\theta})^{\prime}$
and $\hat{\mathcal{I}}=-\sum_{t=1}^{T}h_{t}(\hat{\theta})$. 

The likelihood is based on the assumption that $X_{t}|N_{t}\sim\mathrm{Bin}(N_{t},\lambda_{t})$
and independent across time. Misspecification can arise in two ways.
One way is that the binomial assumption and/or time-independence may
be incorrect (distributional misspecification) and the other is dynamic
misspecification where $\mathbb{E}(X_{t}/N_{t})$ does not agree with
the specified model for $\lambda_{t}$. Under distributional misspecification
the information matrix equality need not hold, and $\hat{\mathcal{J}}$
and $\hat{\mathcal{I}}$ may have different probability limits. If
the independence across $t$ is still valid, such that $s_{t}$ is
not autocorrelated, then we can compute the robust variance covariance
matrix for $\hat{\theta}$, with $\hat{\Sigma}=\hat{\mathcal{I}}^{-1}\hat{\mathcal{J}}\hat{\mathcal{I}}^{-1}$.\footnote{The numerical derivatives computed by the Julia package, $\texttt{ForwardDiff}$,
see \citet{RevelsLubinPapamarkou2016}, are identical to the analytical
expressions for both the score, $s_{t}(\hat{\theta})$ , $t=1,\ldots,T$,
and the hessian, $\sum_{t=1}^{T}h_{t}(\hat{\theta})$.} If the time-independence is also in question, such that the score
may be autocorrelated, we can instead compute: 
\[
\hat{\mathcal{J}}_{K}=\hat{\mathcal{J}}+\sum_{j=1}^{K+1}k(\tfrac{j}{K})\sum_{t=1}^{T-j}\left(s_{t}(\hat{\theta})s_{t+j}(\hat{\theta})^{\prime}+s_{t+j}(\hat{\theta})s_{t}(\hat{\theta})^{\prime}\right),
\]
where $k(x)$ is a kernel function with $k(0)=1$, $k(1)=0$. Dynamic
misspecification is more problematic, as it can result in inconsistent
parameter estimates.
\begin{table}
\caption{Sensitivity of confidence intervals for relative contagiousness.\label{tab:Sensitivity-Robustness}}
\begin{center}
\begin{tabular*}{1\textwidth}{@{\extracolsep{\fill}}ccccc}
\hline 
\noalign{\vskip3pt}
Variance-Estimator &  & Alpha Variant (CI 95\%) &  & Delta Variant (CI 95\%)\tabularnewline[3pt]
\noalign{\vskip3pt}
$\hat{\Sigma}=\hat{\mathcal{I}}^{-1}$ &  & $\gamma_{4.7\mathrm{days}}\in$ {[} 1.5037, 1.5262{]} &  & $\gamma_{4.7\mathrm{days}}\in$ {[} 2.1319, 2.2033{]}\tabularnewline[3pt]
\noalign{\vskip3pt}
$\hat{\Sigma}=\hat{\mathcal{I}}^{-1}\hat{\mathcal{J}}_{0}\hat{\mathcal{I}}^{-1}$ &  & $\gamma_{4.7\mathrm{days}}\in$ {[} 1.4994, 1.5306{]} &  & $\gamma_{4.7\mathrm{days}}\in$ {[} 2.0215, 2.3236{]}\tabularnewline[3pt]
\noalign{\vskip3pt}
$\hat{\Sigma}=\hat{\mathcal{I}}^{-1}\hat{\mathcal{J}}_{1}\hat{\mathcal{I}}^{-1}$ &  & $\gamma_{4.7\mathrm{days}}\in$ {[} 1.4990, 1.5310{]} &  & $\gamma_{4.7\mathrm{days}}\in$ {[} 2.0119, 2.3347{]}\tabularnewline[3pt]
\noalign{\vskip3pt}
$\hat{\Sigma}=\hat{\mathcal{I}}^{-1}\hat{\mathcal{J}}_{2}\hat{\mathcal{I}}^{-1}$ &  & $\gamma_{4.7\mathrm{days}}\in$ {[} 1.4986, 1.5314{]} &  & $\gamma_{4.7\mathrm{days}}\in$ {[} 2.0009, 2.3476{]}\tabularnewline[3pt]
\noalign{\vskip3pt}
$\hat{\Sigma}=\hat{\mathcal{I}}^{-1}\hat{\mathcal{J}}_{3}\hat{\mathcal{I}}^{-1}$ &  & $\gamma_{4.7\mathrm{days}}\in$ {[} 1.4980, 1.5320{]} &  & $\gamma_{4.7\mathrm{days}}\in$ {[} 1.9949, 2.3546{]}\tabularnewline[3pt]
\noalign{\vskip3pt}
$\hat{\Sigma}=\hat{\mathcal{I}}^{-1}\hat{\mathcal{J}}_{4}\hat{\mathcal{I}}^{-1}$ &  & $\gamma_{4.7\mathrm{days}}\in$ {[} 1.4971, 1.5329{]} &  & $\gamma_{4.7\mathrm{days}}\in$ {[} 1.9909, 2.3593{]}\tabularnewline[3pt]
\noalign{\vskip3pt}
$\hat{\Sigma}=\hat{\mathcal{I}}^{-1}\hat{\mathcal{J}}_{5}\hat{\mathcal{I}}^{-1}$ &  & $\gamma_{4.7\mathrm{days}}\in$ {[} 1.4962, 1.5339{]} &  & $\gamma_{4.7\mathrm{days}}\in$ {[} 1.9888, 2.3618{]}\tabularnewline[3pt]
\noalign{\vskip3pt}
$\hat{\Sigma}=\hat{\mathcal{I}}^{-1}\hat{\mathcal{J}}_{6}\hat{\mathcal{I}}^{-1}$ &  & $\gamma_{4.7\mathrm{days}}\in$ {[} 1.4952, 1.5349{]} &  & $\gamma_{4.7\mathrm{days}}\in$ {[} 1.9888, 2.3618{]}\tabularnewline[3pt]
\hline 
\end{tabular*}
\end{center}
     \footnotesize
     \renewcommand{\baselineskip}{11pt}
     \textbf{Note:} Sensitivity of confidence intervals for relative contagiousness,
    $\gamma_{4.7\mathrm{days}}$, to the choice of variance-covariance estimator
    (computed with non-robust and various robust standard errors).
\end{table}

Our empirical results are based on the autocorrelation robust estimator,
$\hat{\Sigma}=\hat{\mathcal{I}}^{-1}\hat{\mathcal{J}}_{K}\hat{\mathcal{I}}^{-1}$,
with $K=4$ and the Parzen kernel function for $k(\cdot)$. Standard
errors and confidence intervals for $\alpha$ and $\beta$ are given
from the diagonal elements of $\hat{\Sigma}$, which we denote by
$\hat{\sigma}_{\alpha}^{2}$ and $\hat{\sigma}_{\beta}^{2}$, respectively.
The reported 95\% confidence intervals for $\alpha$ and $\beta$
are based on the point estimates $\pm1.96$ times the corresponding
standard error. Those for $\gamma_{4.7\mathrm{days}}=\exp(\tfrac{4.7}{7}\beta)$
are given by $\exp\{\tfrac{4.7}{7}(\hat{\beta}\pm1.96\hat{\sigma}_{\beta})\}$.

\end{document}